\documentclass[twocolumn,showpacs,preprintnumbers,amsmath,amssymb]{revtex4}

\usepackage{graphicx}
\usepackage{dcolumn}
\usepackage{bm}

\begin{document}

\vspace{5mm}

\newcommand{\goo}{\,\raisebox{-.5ex}{$\stackrel{>}{\scriptstyle\sim}$}\,}
\newcommand{\loo}{\,\raisebox{-.5ex}{$\stackrel{<}{\scriptstyle\sim}$}\,}

\title{Coupling dynamical and statistical mechanisms for baryonic cluster 
production in nucleus collisions of intermediate and high energies.} 

\author{A.S.~Botvina$^{1,2,3}$, N.~Buyukcizmeci$^{4}$, M.~Bleicher$^{1,3,5,6}$}

\affiliation {$^1$ITP J.W. Goethe 
University, D-60438 Frankfurt am Main, Germany} 

\affiliation {$^2$Institute for Nuclear 
Research, Russian Academy of Sciences, 117312 Moscow, Russia} 

\affiliation{$^3$ Helmholtz Research Academy Hesse for FAIR (HFHF), 
GSI Helmholtz Center, Campus Frankfurt, Max-von-Laue-Str. 12, 
60438 Frankfurt am Main, Germany}

\affiliation{$^4$Department of Physics, Selcuk University, 42079 Kampus, 
Konya, Turkey}

\affiliation {$^5$GSI Helmholtz Center for Heavy Ion Research, 
Planckstr.1, Darmstadt, Germany} 

\affiliation {$^6$John-von-Neumann Institute for Computing (NIC), 
FZ J\"ulich, J\"ulich, Germany}

\date{\today}

\begin{abstract}

Central nucleus-nucleus collisions produce many new baryons and the 
nuclear clusters can be formed from these species. 
The phenomenological coalescence models were sufficiently good for 
description of light nuclei yields in a very broad range of collision 
energies. We demonstrate that in reality the coalescence process 
can be considered as 1) the formation of primary diluted excited baryon 
clusters and 2) their following statistical decay leading to the final 
cold fragment production. We argue that the formation of such excited 
systems from the interacting baryons is a natural consequence of the 
nuclear interaction at subnuclear densities resulting 
in the nuclear liquid-gas type phase transition in finite systems. In this 
way one can provide a consistent interpretation of the experimental 
fragment yields (FOPI data) including the important collision energy 
dependence in relativistic ion reactions. We investigate the regularities 
of this new kind of fragment production, for example, their yield, 
isospin, and kinetic energy characteristics. 
A generalization of such a clusterization mechanism for hypernuclear matter 
is suggested. The isotope yields and particle 
correlations should be adequate for studying these phenomena. 

\end{abstract}

\pacs{25.75.-q , 21.80.+a , 25.70.Mn }

\maketitle
 
\section{Introduction}

The production of nuclear fragments in relativistic nuclear reactions is 
one of the important topics in nuclear physics. 
It is known since the late 1970s that many different light 
complex nuclei can be produced in central nucleus-nucleus collisions 
\cite{Gos77}. Usually it is associated with a coalescence-like mechanism, 
i.e., the complex particle are formed from the dynamically produced nucleons 
and other baryons, because of their attractive interaction. The coalescence 
model has demonstrated a good description of the data (by adjusting the 
coalescence parameter) from intermediate to very high collision energies 
\cite{Ton83,Bot17a,Neu03}. There were many other intensive 
investigations of the coalescence mechanism, e.g., see the latest 
Refs. \cite{Sch99,Som19,Blu19}. 
This supports the idea that the baryons emerging 
after the initial dynamical stage are the main constituents of
these nuclei.

On the other hand, many nuclear fragments can be produced in peripheral 
collisions as a result of multifragmentation of hot projectile/target-like 
residual nuclei. A lot of experiments was devoted to this study 
associated with the nuclear liquid-gas type phase transition. 
In particular, ALADIN \cite{ALADIN,Bot95,Xi97,Ogu11}, EOS \cite{EOS}, 
ISIS \cite{Vio01,Pie02}, FASA \cite{FASA}, 
and other experimental collaborations have 
provided very high quality data. From theoretical side many 
dynamical and statistical models were developed. Here we remind on 
success of the hybrid approaches which include the descriptions of 
the non-equilibrium dynamical reaction stage, and the following 
decay of the equilibrated nuclear sources. The description of the last 
stage with the statistical models was very instructive (see, e.g., 
SMM \cite{SMM} and MMMC \cite{MMMC}). These statistical models provide the 
generalization of the liquid-gas type phase transition phenomenon to finite 
nuclear systems. 
The success of the statistical models in description of the fragment 
production has encourage to generalize them for hypernuclear matter, 
and, finally, for production of hypernuclei \cite{Bot07}. The involvement of 
hyperons ($\Lambda$, $\Sigma$, $\Xi$, $\Omega$) obtained in high-energy reactions 
provides a complementary method to improve traditional nuclear studies 
and opens new horizons for studying particle physics and nuclear 
astrophysics (see, e.g., \cite{Sch93,Has06,Gal12,Buy13,Hel14} 
and references therein). 
Previously we have theoretically investigated the production regularities 
of large hypernuclei which can originate from peripheral nucleus-nucleus 
collisions. In this case the produced strange particles are captured by the 
projectile and target residues. 
In particular, we have demonstrated a big yield of 
such hypernuclei, their broad distribution in mass and isospin, and 
a considerable production of multi-strange hypernuclei 
\cite{Bot11,Bot13,Bot16,Bot17}. This opens 
new possibilities for their investigation in comparison with traditional 
hypernuclei experiments with light particles. 

However, many modern experimental detectors for heavy-ion collisions 
are designed to measure particles produced in midrapidity reaction zone. 
Presently some experimental collaborations (STAR at RHIC \cite{star}, 
ALICE at LHC \cite{alice}, CBM \cite{Vas17},  BM@N, MPD 
at NICA \cite{nica}) plan to investigate light nuclei clusters and 
their properties in reactions induced by relativistic hadrons and ions. 
Therefore, in this paper we concentrate on the production of light nuclei (and 
hypernuclei) coming from central high-energy collisions. 

To understand these nuclei formation process we apply theoretical methods, 
which were partly developed earlier, e.g., for the coalescence 
procedure \cite{Bot15}. In this paper we emphasize another aspect of 
this phenomenon: The coalescent clusters can be sufficiently large and present 
pieces of nuclear matter at subnuclear densities. They can have some 
excitation energy and their following evolution can be described by the 
statistical methods. Moreover, the decay of these excited clusters can be 
treated by basing on the previous theoretical and experimental achievements 
concerning nuclear multifragmentation phenomena. Contrary to the standard 
coalescence picture which considers only baryons combining into a final 
nucleus the new picture includes effectively many body interaction, also 
with baryons which were not captured in the final nuclei. This novel 
development leads to the qualitatively new predictions and describe 
experimental data which were never analyzed before. 
Below we demonstrate the important new findings and compare our 
results with recent FOPI experimental data \cite{FOPI1997,FOPI2010}. 

\section{Physical meaning of the coalescence into hot clusters} 

The normal and strange baryons are abundantly produced in high energy 
particle reactions, e.g., nucleus-nucleus, hadron-nucleus and lepton-nucleus 
collisions. For description of this process one can use the transport 
models, like UrQMD \cite{urqmd1,urqmd2}, HSD \cite{HSD}, IQMD \cite{IQMD}, 
GiBUU \cite{giessen}, and others. 
These models generate baryons 
coming from primary and secondary particle interactions, including the 
rescattering and decay of resonances. 
In the end of the dynamical stage these produced baryons can also attract 
each other and form clusters. The phenomenological coalescence models are 
usually adjusted to describe the cluster yield by using a 
coalescence parameter. The success in description of the experimental 
data (see, e.g., \cite{Ton83,Bot17a,Neu03,Sch99,Som19,Blu19} and references 
in) tells us that the clusters can really be 
consisted of the dynamical baryons. However, it does not tell us about 
the phase space distribution of the involved baryons, 
and on properties of the formed clusters. As a rule 
the transport models designed to describe high energy interactions have no 
possibilities to follow precisely low energy interactions between nucleons 
leading to the nucleus formation at low densities. 

In the end of the dynamical stage (at time around $\sim$10-30 fm/c after 
the beginning of the nucleus collision) some produced baryons can be 
located in the vicinity of each other with local subnuclear 
densities around $\sim$0.1$\rho_0$ 
($\rho_0 \approx 0.15$ fm$^{-3}$ is the normal nuclear density). The momenta 
of these baryons are obtained from primary nucleons and other hadrons 
interactions during their collisions. These particles are mostly 
concentrated in the midrapidity region. We expect that by this time 
the fast produced particles, as well as the nucleons of projectile 
and target residue, have separated sufficiently, so the hard interactions 
leading to the new particle formation are practically stopped. Such kind a 
saturation is demonstrated in many transport approaches \cite{Bot11}. 
For this reason one can also expect that if some baryonic clusters are possibly 
formed via dynamical correlations in earlier times they will be destroyed 
by intensive interactions existing at large densities. At the subnuclear 
density the baryons in the coordinate vicinity will still have an attractive 
nuclear interaction and may form new baryon clusters. 
Since the baryons can move respect to each other inside these clusters, 
we may say we are dealing with the excited 
clusters. The new idea is that our coalescence procedure can be presented 
as a division of the all phase space into small parts (clusters) with baryons which are 
in equilibrium respective to the nucleation process.  
The following evolution of such clusters, 
including the formation of nuclei from these baryons, can be described in 
the statistical way. In other words, these hot clusters decay into nuclei. 
We emphasize a very important difference of our mechanism from the standard 
coalescence: It is assumed in the simplistic coalescence picture 
that only baryons which combine a nucleus can interact in the final state. 
All other baryons will not interact with this nucleus, or interact 
very slightly by taking extra energy to conserve the momentum/energy balance. 
In our case all baryons of the primary hot coalescent cluster interact intensively 
to produce final nuclei. Even though not all these baryons will be bound in 
the nuclei in the end. 

Here the crucial point is if the lifetime of these clusters is sufficient 
for equilibration between the baryons to be considered as statistical 
systems and to apply the statistical methods. 
We remind that the lifetime of finite nuclear species is related to the 
energy implemented into these species. We know from the extensive studies of 
nuclear multifragmentation reactions \cite{Bot95,Xi97,Ogu11,SMM,MMMC} 
that the excitation energies 
of the excited nuclear systems can reach up to 8--10 MeV per nucleon, and 
the statistical models describe their disintegration very good. We've also 
learned from the analysis of nuclei production in multifragmentation that the 
densities before the break-up of these systems are around 0.1--0.3$\rho_0$, 
and their lifetime is 50--100 fm/c \cite{Vio01,FASA}. 
We believe that the difference between  the multifragmentation of excited 
projectile- and target-like sources and formation of the baryon clusters in 
the midrapidity zone is just in the dynamical mechanisms leading to these 
finite systems. In the standard 
multifragmentation the systems are prepared via dynamical knocked many 
nucleons and thermal (or dynamical) expansion of the remaining diluted 
nuclei. Our cluster systems are prepared just as a result of the local 
interaction (e.g., attraction) of the stochastically produced primary 
baryons. Therefore, we can suggest that the energy around 
$\sim$10 MeV per nucleon could be a reasonable value 
which can be reached in such hot coalescent clusters, similar to the 
standard multifragmentation case. If the excitation energy is much higher, 
then the existence of such clusters as intermediate finite systems with their 
following evolution in the statistical way become problematic. 

\section{Simulations of produced baryons and their coalescence} 

It is natural to use the transport models for the generation 
of baryon parameters (coordinates and momenta) after their dynamical 
production in relativistic nucleus collisions. As a first approximation 
we can employ the Dubna Cascade model (DCM) which has demonstrated 
a good performance in description of many experimental data 
\cite{Ton83,Bot11,Bot17}. 
According to the construction this model provides a defined time-end of the 
fast reaction stage and gives the corresponding parameters of baryons. 
In the following we note it as G1 generation. 
However, in order to understand the coalescence and following 
de-excitation processes better, as an 
initial step we'll use the distributions of baryons in kinetic energy 
obtained within other models which have clear physical interpretation. 
Our second method is noted as G2 generation: We perform the isotropic 
generation of all baryons of the excited sources according the microcanonical 
momentum phase space distribution with the total momentum and energy 
conservation. It is assumed that all particles are in a large freeze-out 
volume (at subnuclear densities) where they can still interact to populate 
uniformly the phase space.  
Technically, it is done  with the Monte-Carlo 
method applied previously in the SMM and Fermi-break-up model 
in the microcanonical way \cite{SMM}, 
and taking into account the relativistic effects according to the 
relativistic connection between momentum $\vec{p}$, mass $m$, and kinetic 
energy of particles $E_{0}$, see eq.~(\ref{relc}) (where the sum is over all 
particles and all ingredients are taken in the energy units). 
\begin{equation} \label{relc}
\sum \sqrt {\vec{p}^2 + m^2} = E_0 + \sum m .
\end{equation}
The total energy available for kinetic motion of baryons 
$E_{0}$ (we call it as the source energy) is the important parameter 
which can be adjusted to describe the energy introduced into the system 
after the dynamical stage. We believe G2 generation can be considered 
as one of the reasonable cases since there are very intensive interactions 
between colliding nucleons of target and projectile, which take place in 
some extended volume during the reaction and may lead to the equilibration 
in the one-particle degrees of freedom. In this case we do not take directly 
into account the coordinates of the baryons but we assume they are 
proportional to their velocities and strictly correlate with them. 
\begin{figure}[tbh]
\includegraphics[width=8.5cm,height=13cm]{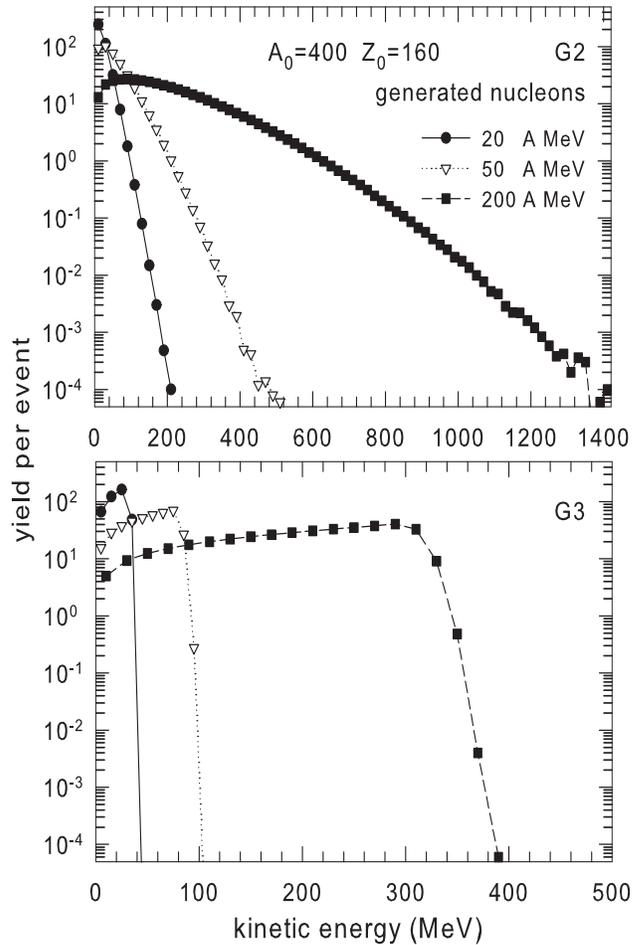}
\caption{\small{ 
Energy spectra for initial nucleons of the hot expanding nuclear system 
according to the microcanonical phase space distribution - G2 (top panel), 
and according to the hydrodynamical-like explosion - G3 (bottom panel). 
The suggested total kinetic energies are 20, 50, and 200 MeV per nucleon. 
The nucleon source size and composition are shown in the top panel.
}}
\label{fig1}
\end{figure}

\begin{figure}[tbh]
\includegraphics[width=9cm,height=16cm]{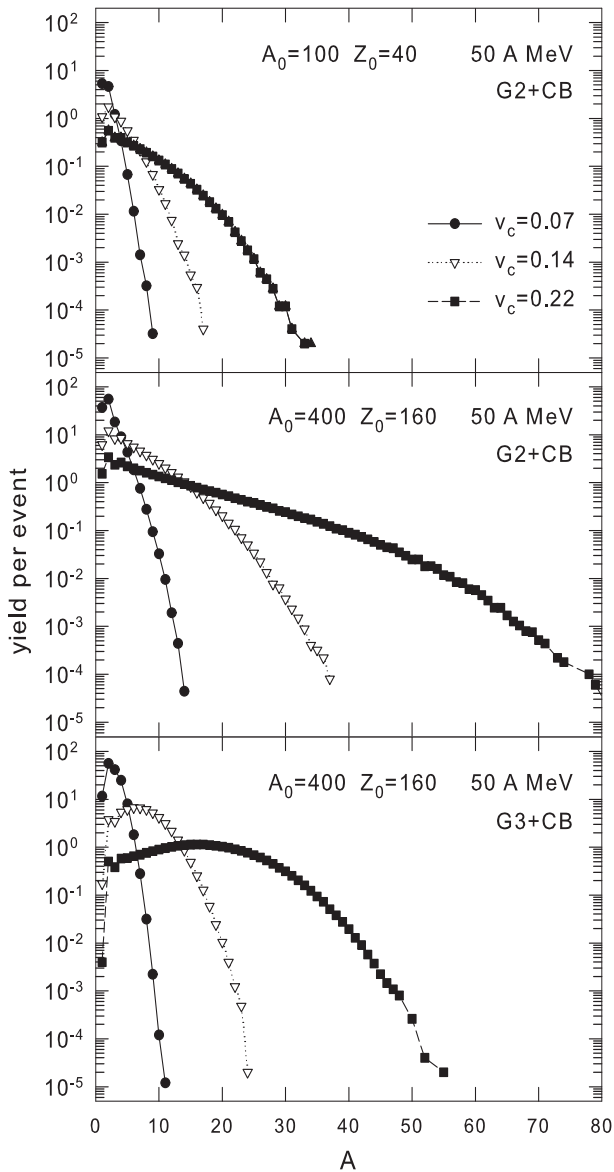}
\caption{\small{ 
Yield of coalescent clusters versus their mass number A after the CB 
calculations at the source energy of 50 MeV per nucleon. Composition and 
sizes of sources, nucleon generators (G2 and G3), as well as coalescence 
parameters ($v_c$) are indicated in the panels. 
}}
\label{fig2}
\end{figure}
In the third method, G3 generator, we assume the momentum generation 
similar to the explosive 
hydrodynamical process when all nucleons fly out from the center of the 
system with the velocities exactly proportional to their coordinate distance 
to the center of mass.  For this purpose, with the Monte-Carlo method, we 
place uniformly all nucleons inside the sphere with the radius 
$F\cdot R_n \cdot A_0^{1/3}$ without overlapping. 
Here $A_0$ is the nucleon number, and $R_n \approx 1.2$~fm is the nucleon 
radius. The size factor $F \approx 3$ is assumed for the expanded freeze-out 
volume where the nucleon can still strongly interact with each other. 
At the intermediate collision energies this volume corresponds approximately 
to the average expansion of the system after simulations with the 
transport models, when the baryon interaction rate drastically decreases. 
Finally, we attribute 
to each nucleon the velocity by taking into account the momentum and energy 
conservation for the relativistic case (eq.~\ref{relc}). 
Obviously, the velocities and coordinates of baryons are strongly 
correlated with each other. We think it can also be considered as another 
case important for our study. 

In the top part of Fig.~\ref{fig1} 
we demonstrate the energy distribution of all initial nucleons in the 
excited source with mass number $A_0$=400, charge $Z_0$=160, after G2 
generation. The source energy (i.e., the total kinetic energies of all 
nucleons) were taken as 
$E_{0}$=20, 50, and 200 A MeV. This may characterize the hot systems 
produced at central collisions of heavy nuclei at laboratory energies 
around 100--1000 A MeV. As expected, the distributions are very broad. 
We have also checked that the size effect on the distributions is 
practically minimal, as it is following thermodynamical quantities in the 
one-particle approximation. 
In the bottom part we show the same distributions but for G3 generation. 
It is seen a qualitative difference of the nucleon energy distributions 
after G2 and G3 generators. 
G3 provides a very compact distribution of nucleons according to their 
positions in the freeze-out volume. We think it is important to demonstrate 
how this difference will be manifested in the cluster production and the 
kinetic energy of clusters. 

As discussed, the final subtle interactions (attraction) of the 
generated baryons 
can lead to the cluster formation. To describe it we use the coalescence 
prescription, and apply the coalescence of baryon (CB) model 
\cite{neu00,Bot15}. In G2 and G3 cases the criterion is the proximity of the 
velocities (or momenta) of nucleons. As was mentioned, in these cases we do 
not include 
explicitly the coordinate of nucleons, since this kind of generation 
suggests a correlation of velocities and space coordinates. In particular, 
the coordinate vectors should be directly proportional to the velocities 
vectors. So the velocity coalescence parameter is sufficient for the 
cluster identification in these models. Such a correlation exists in 
many explosive processes and it influences the original clusterization. 
However, for the following evaluation of the cluster properties we assume 
that such clusters with nucleons inside have the density of 
$\rho_c \approx \frac{1}{6} \rho_0$ as it was established in the previous 
studies of statistical multifragmentation process \cite{SMM,MMMC,Vio01,FASA}. 
This corresponds to the average distance of around 2 fm between neighbour 
nucleons, and these nucleons can still interact leading to the nuclei 
formation. It is also consistent with the densities which can be obtained 
for such clusters with the transport model calculations in the end of the 
dynamical stage (see Section 5). 
Within the CB model we suggest that baryons (both nucleons and hyperons) can 
produce a cluster with mass number $A$ if their velocities relative to the 
center-of-mass velocity of the cluster is less than $v_c$. Accordingly we 
require $|\vec{v}_{i}-\vec{v}_{cm}|<v_{c}$ for all $i=1,...,A$, where 
$\vec{v}_{cm}=\frac{1}{E_A}\sum_{i=1}^{A}\vec{p}_{i}$ ($\vec{p}_{i}$ are 
momenta and $E_A$ is the sum energy of the baryons in the cluster). 
This is performed by sequential comparison of the velocities of all baryons. 
As done before \cite{neu00,Bot15}, to avoid the problem related to the 
sequence of nucleons within the algorithm, we apply the iterative coalescence 
procedure, starting from the diminished coalescence parameters for clusters 
and by increasing them step-by-step up to the $v_c$ value. 

We show in Fig.~\ref{fig2} the distributions of clusters in their mass number 
$A$ after the coalescence of initial nucleons of the primary source 
$A_0$=100, $Z_0$=40 (top panel), and $A_0$=400, $Z_0$=160 (middle and 
bottom panels), for $E_{0}$=50 A MeV, for the velocity coalescence parameter 
$v_c$=0.07, 0.14, and 0.22 $c$. In our case $v_c$ means the maximum velocity 
deviation and all baryons with lower relative velocities do compose a cluster. 
The largest $v_c$=0.22 $c$ is approximately of the order of 
the Fermi-velocity which is expected in such nuclei. The smallest 
$v_c$=0.07 $c$ is consistent with the coalescence parameters extracted 
previously in analyses of experimental data \cite{Ton83,Bot17a}. 
\begin{figure}[tbh]
\includegraphics[width=8.5cm,height=13cm]{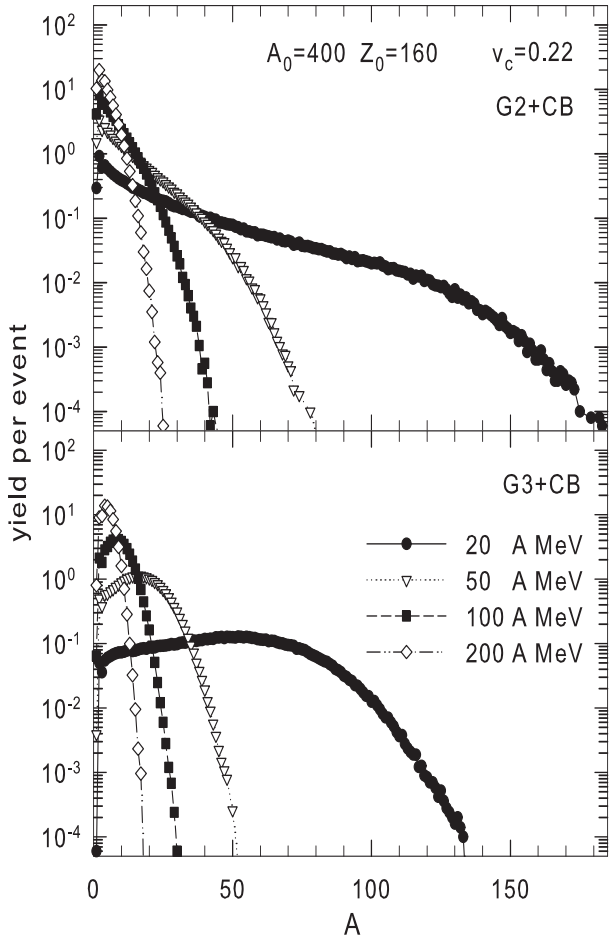}
\caption{\small{ 
The same as in Fig.~\ref{fig2} but for fixed coalescence parameter 
$v_c$=0.22 $c$, G2 and G3 generators. The source composition and 
energies are shown in the panels. 
}}
\label{fig3}
\end{figure}
In the latest case the cluster excitation energy is minimal and the cluster may 
not decay by the nucleon emission. 

One can see that the big clusters indeed can be produced with the 
coalescence mechanism. It was discussed previously \cite{neu00,Bot15}, 
however, without determining the cluster properties. 
By comparing the middle and top panels it is instructive to note 
that the bigger source can produce larger 
clusters at the same initial excitation per nucleon. This is a typical 
collective effect coming from the larger number of nucleons involved in 
the reaction. It is obvious that a larger coalescence parameter leads to 
the formation of bigger clusters. Still, as will be shown below, their 
excitation energies will be also higher, and their subsequent decay 
decreases the nuclei sizes. 
By comparing the results after G2 and G3 generators (middle and 
bottom panels) one can see an essential difference in the produced clusters.
In the last case the cluster have large sizes which are likely grouped 
around the mean values with the maximum yield. 
This is the consequence of flow-like initial distribution of baryons. 
\begin{figure}[tbh]
\includegraphics[width=8.5cm,height=13cm]{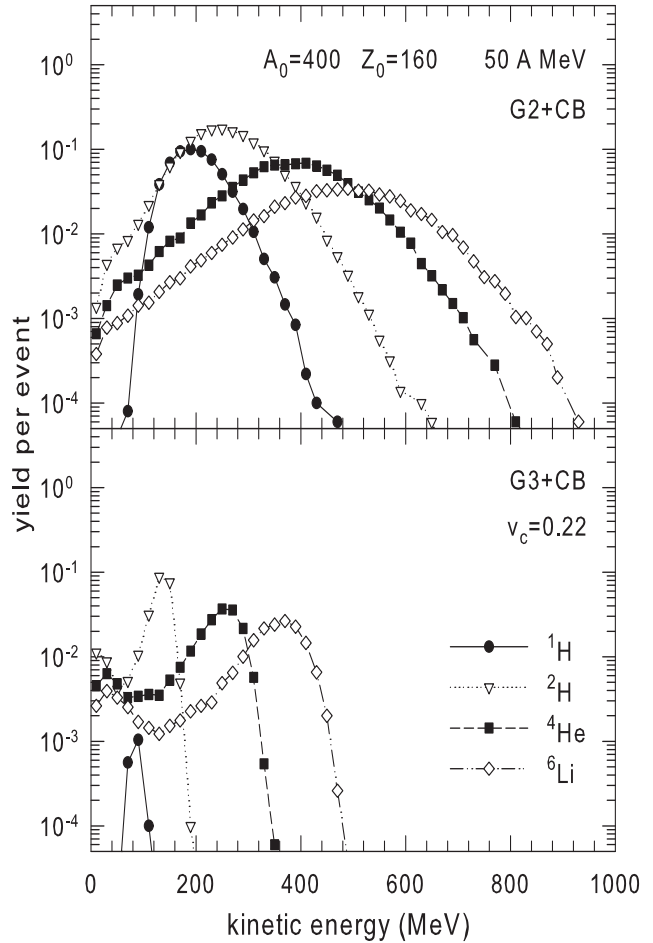}
\caption{\small{ 
Energy distributions of protons and some light particles after the 
coalescence. The source characteristics, baryon generators, coalescence 
parameter, and produced species are indicated in the panels. 
}}
\label{fig4}
\end{figure}
On the other hand  in G2 case we 
can get very big clusters, however, with a low probability. They come from 
the low-energy component of the G2 nucleon energy distribution. 
\begin{figure}[tbh]
\includegraphics[width=8.5cm,height=13cm]{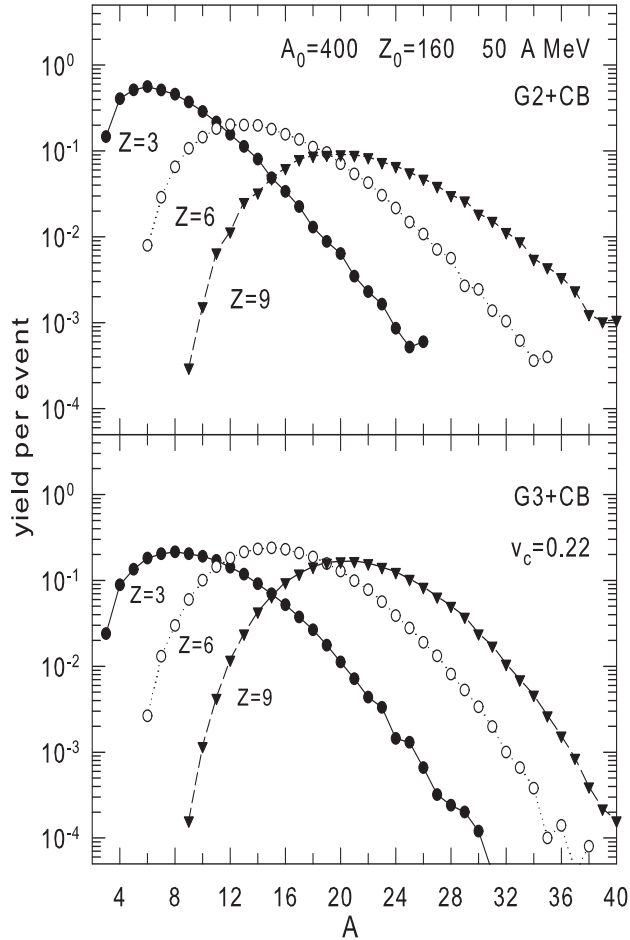}
\caption{\small{ 
Isotope distributions of elements with charges $Z$=3, 6 and 9 after the 
coalescence in the sources after G2 (top panel) G3 (bottom panel) baryon 
generators. 
The source sizes, energies and coalescence parameter are shown in the figure. 
}}
\label{fig5}
\end{figure}
\begin{figure}[tbh]
\includegraphics[width=8.5cm,height=13cm]{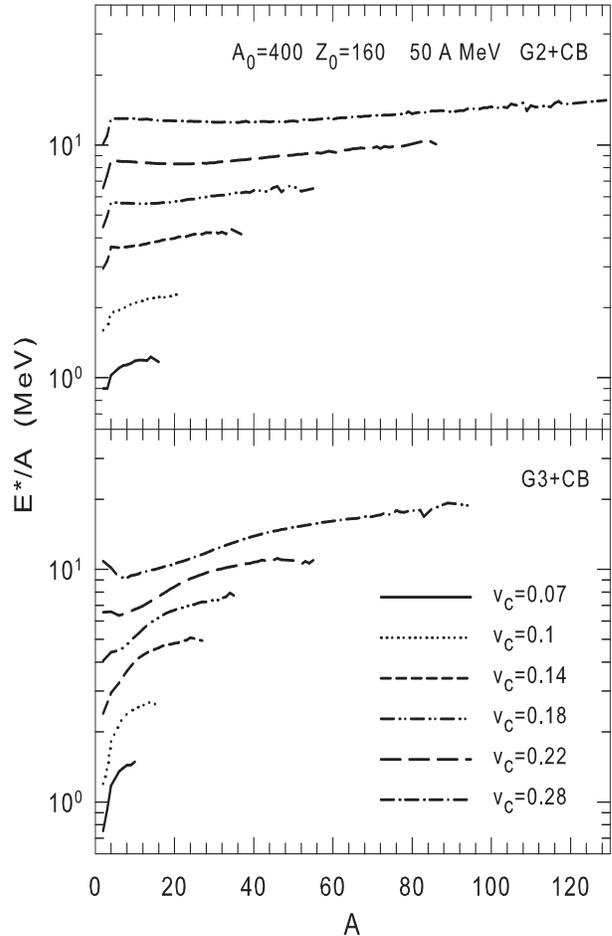}
\caption{\small{ 
Average internal excitation energy of coalescent clusters versus their mass 
number $A$ produced as a result of 
the coalescence (CB) in the sources with $A_0$=400 and $Z_0$=160 
after G2 (top panel) and G3 (bottom panel). 
The source energy and coalescence parameters are shown in the figure. 
}}
\label{fig6}
\end{figure}

It is important to understand how the masses of coalescent 
clusters evolve with 
the source energy. In Fig.~\ref{fig3} we demonstrate the mass distributions 
in the biggest sources at the parameter $v_c$=0.22 $c$ for the wide range of 
$E_{0}$. The yields of big clusters are larger at the low source energy, 
since the velocities of nucleons are smaller and closer to each other to 
form a cluster. However, there are a lot of intermediate mass clusters 
(with $A \goo$10) even 
at high source energies. It is a consequence of the stochastic nature for 
production of such nucleons since they may appear in the phase space vicinity 
of other nucleons. 
Under the assumptions of G2 and G3 generators we simulate it by the 
Monte-Carlo method. The considered source energies correspond to nucleons 
originated from central heavy ion collisions with beam energies less 
than 1~A~GeV. 

The kinetic energy of produced clusters is also an important characteristic 
which can give experimental evidences about baryons composing clusters. 
Fig.~\ref{fig4} 
demonstrates the kinetic energies of the remaining protons, and clusters 
$^2$H, $^4$He and $^6$Li after the coalescence 
in the $A_0$=400, $Z_0$=160, $E_{0}$=50 A MeV source, for $v_c$=0.22 $c$. 
It is clearly seen that, for example, the spectrum for remaining protons is 
essentially different from the initial distributions of nucleons shown in 
Fig.~\ref{fig1}: The reason is that a lot of protons are captured by primary 
clusters. In G3 case practically all protons are in coalescent clusters. 
After decay of the excited clusters many protons may become 
free again (see Fig.~10), however, this additional interaction 
via the clusterization may change their energy spectra. 
One can see also that the energy distributions of the produced 
clusters have a flow-like structure, i.e., each captured nucleon adds 
the kinetic energy to the cluster. It is especially seen in G3 case, 
where the kinetic energy of clusters with the maximum yield is nearly 
directly proportional to their mass number. 

In addition, these clusters can have very exotic isospin composition. This 
is a direct consequence of the initial random  distribution of protons and 
neutrons in the phase space. In Fig.~\ref{fig5} 
we demonstrate such broad isotope distributions 
for few elements. It is clear from general properties of nuclei that these 
isotopes can not be stable and must decay afterwards. However, this decay 
will take 
place during the time which is more prolonged than the dynamical reaction 
stage. As a result of the secondary processes we can expect very exotic 
nuclear species in these reactions. 
We have found also that the bigger source can be responsible for larger 
neutron 
enrichment and, consequently, more exotic nuclei. 

As well known the nuclei have many excited states which decay during the time 
much longer than the dynamical (collision) reaction time which is around few 
tens fm/c. Generally, we expect that during the coalescence process the 
highly excited coalescence clusters can be produced. Moreover, as we have 
discussed in Section~2, the subtle interaction of dynamically produced 
baryons can result into excited systems which decay later on in a 
statistical way. In the lowest limit we 
can estimate this excitation as a relative motion of the nucleons initially 
captured into a cluster respective to the center of mass of this cluster. 
In this case the excitation energy $E^{*}$ of the clusters 
with mass number $A$ and charge $Z$ is calculated as 
\begin{equation} \label{excit}
E^{*}= \sum_{i=1}^{A}\sqrt{\vec{p}_{r i}^{ 2}+m_{i}^{2}} - M_{A} ,
\end{equation}
where $M_{A}$ is the sum mass of nucleons in this nuclear cluster, 
$i=1,...,A$ enumerate nucleons in the cluster, 
$m_{i}$ are the masses of the individual nucleons in the cluster, 
$\vec{p}_{r i}$
are their relative momenta (respective to the center mass of the cluster). 
However, in the cluster volume the nucleons can interact with each other 
and the binding interaction energy $\delta E^{*}$ should be added to 
the $E^{*}$. As an upper limit we can take the ground state binding energy 
of normal nuclei with $A$ and $Z$. However, 
we believe that in fact this energy should be lower since the nuclear 
cluster is already expanded as a result of the dynamical reaction stage. 
Therefore, as the first approximation we use the following recipe for 
evaluation of $\delta E^{*}$: It is 
known the ground state binding energy of nuclei can be written as the 
sum of short range contributions ($E_{sr}$, which naturally includes volume, 
symmetry, surface energies), and the long-range Coulomb energy ($E_{col}$), 
see, e.g., Ref.~\cite{SMM}. Since a cluster is extended its Coulomb 
energy contribution will be smaller and we can recalculate it proportional 
to $\left(\frac{\rho_c}{\rho_0}\right)^{1/3}$ (in the Wigner-Seitz 
approximation \cite{SMM}). For the short range energies, 
it is assumed that all contributions do also decreases proportional to 
$\left( \frac{\rho_c}{\rho_0} \right)^{2/3}$ as it follows 
from the decreasing of the Fermi energy of nuclear systems. In future we plan 
to investigate the excitation energy problem in details. In this 
work we use 
\begin{equation} \label{deltaE}
\delta E^{*}= E_{col}\left(\frac{\rho_c}{\rho_0}\right)^{1/3} + 
E_{sr}\left(\frac{\rho_c}{\rho_0}\right)^{2/3}~, 
\end{equation}
since it provides a reasonable estimate in between the two limits. 

As usual we consider the cases of both G2 and G3 baryon generators. 
In Fig.~\ref{fig6} 
we present the average excitation energies of such clusters versus their 
mass number for the big systems $A_0$=400, $Z_0$=160, and $E_{0}$=50 A MeV, with the 
coalescence parameters $v_c$ from 0.07, to 0.28 $c$. One can see that 
the excitation energy per nucleon increases with this parameter. This is 
because more nucleons with large relative velocities are captured into the 
same cluster. By comparing the panels of Fig.~\ref{fig6} we see the effect 
of the source generator on these distributions: The excitations are not very 
different, since 
they are determined by relative nucleon motions inside clusters. 
Nevertheless the G3 provides a general increase of the excitation with 
the mass number since the large clusters are consisting of baryons 
having initially higher velocities. 

\section{Disintegration of hot coalescent clusters} 

It is clear that our primary coalescent nuclear clusters must disintegrate 
into small peaces because of their big excitation energy. 
As we discussed in section~2 the whole process of both the formation and 
subsequent decay of such clusters is the necessary part of one physical 
phenomenon. It can be considered as a result of the 
residual nuclear interaction between baryons at the subnuclear density 
leading to the production of final nuclear species. 
In the end the cold and stable nuclei are produced. 
At this point it is instructive to recall 
the previous analyses of experimental data on disintegration of excited 
nuclear systems \cite{ALADIN,Bot95,Xi97,Ogu11,EOS,Vio01,Pie02,FASA,SMM,MMMC}. 
This investigation has lead to the conclusions on the 
statistical nature of such disintegration. Also it was discussed that this 
process can be the manifestation of the liquid-gas type phase transition in 
finite nuclei systems \cite{SMM}. 
We remind, it was obtained in these theoretical analyses 
\cite{Bot95,Xi97,Ogu11,MMMC} that there is 
the limitation for the excitation energy for the finite thermalized nuclear 
systems, around 10 MeV per nucleon, with values closed to the binding 
energies of the systems. As was established these systems decay in time 
about $\sim$100 fm/c \cite{Vio01,Pie02,FASA} 
that is several times longer than the dynamical 
reaction stage. This result is obtained in multifragmentation 
of nuclear residues produced in peripheral relativistic ion collisions. 
We believe that it is a general property of finite nuclear systems: 
Independent on the way how the primary excited clusters are formed, they 
can be considered as small systems of interacting nucleons in the region 
of the nuclear-liquid gas coexistence. As a result we can also expect 
the same limitation in the excitation energy of our coalescent clusters. 
This puts natural limits 
on the values of the parameters $v_c$ for the coalescence mechanism in 
central nuclei collisions. In the following we can apply the well 
established statistical models for the cluster de-excitation. 
\begin{figure}[tbh]
\includegraphics[width=8.5cm,height=13cm]{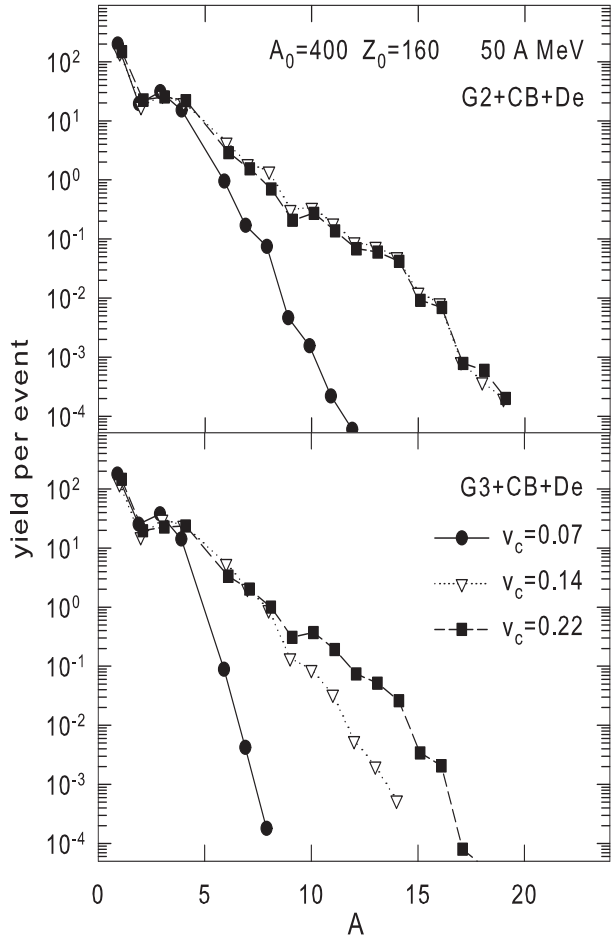}
\caption{\small{ 
Yield of final cold fragments versus their mass number A after the 
coalescence and fragment de-excitation (CB+De) 
calculations at the source energy of 50 MeV per nucleon. Baryon generators, 
composition and 
sizes of sources, as well as coalescence parameters are shown in the panels. 
}}
\label{fig7}
\end{figure}
\begin{figure}[tbh]
\includegraphics[width=8.5cm,height=13cm]{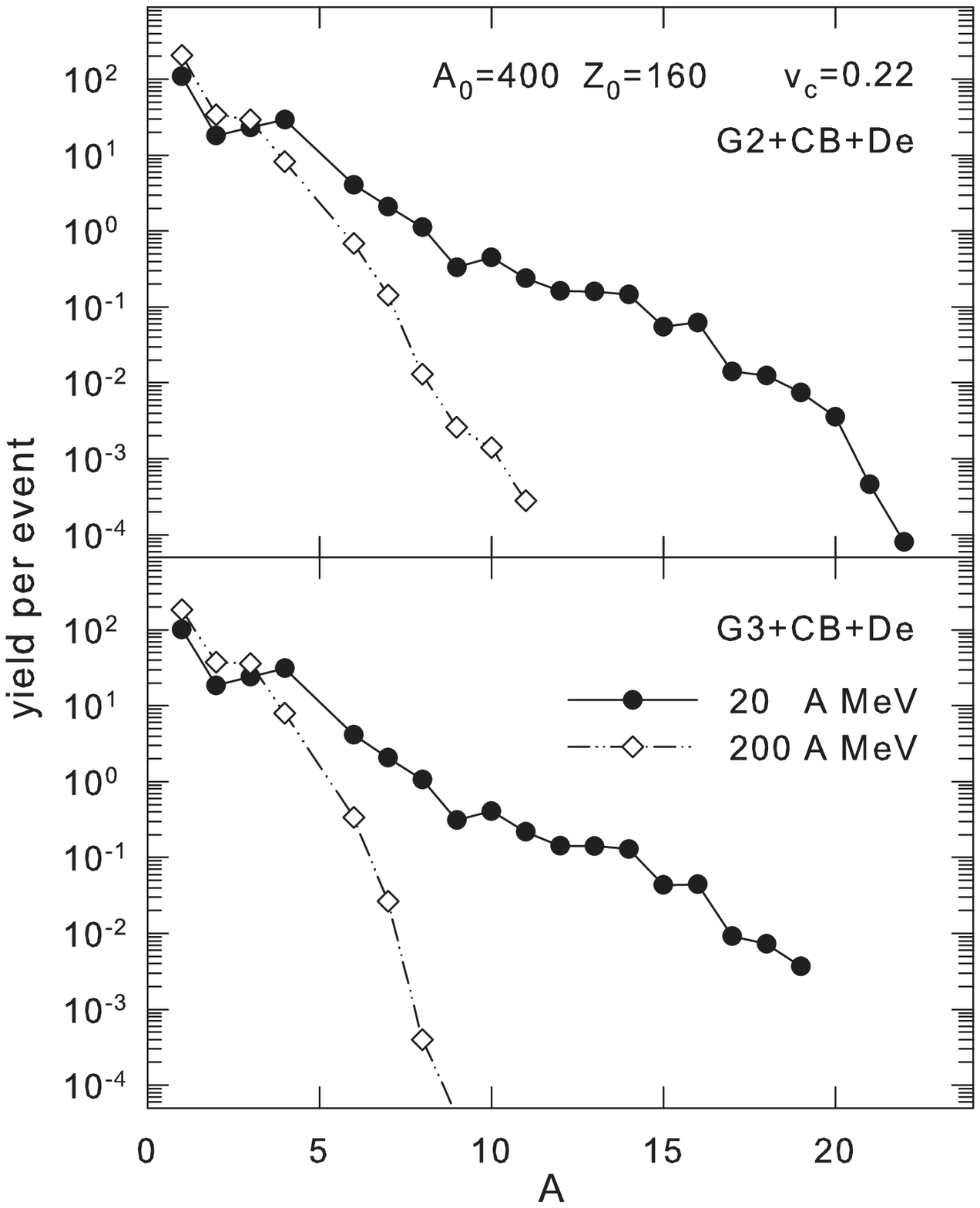}
\caption{\small{ 
Yield of final cold fragments versus their mass number A after the 
coalescence and fragment de-excitation (CB+De) 
calculations at the source energies of 20 and 200 A MeV per nucleon. 
Baryon generators, composition and 
size of sources, as well as the coalescence parameter are shown in the panels. 
}}
\label{fig8}
\end{figure}

\begin{figure}[tbh]
\includegraphics[width=8.5cm,height=13cm]{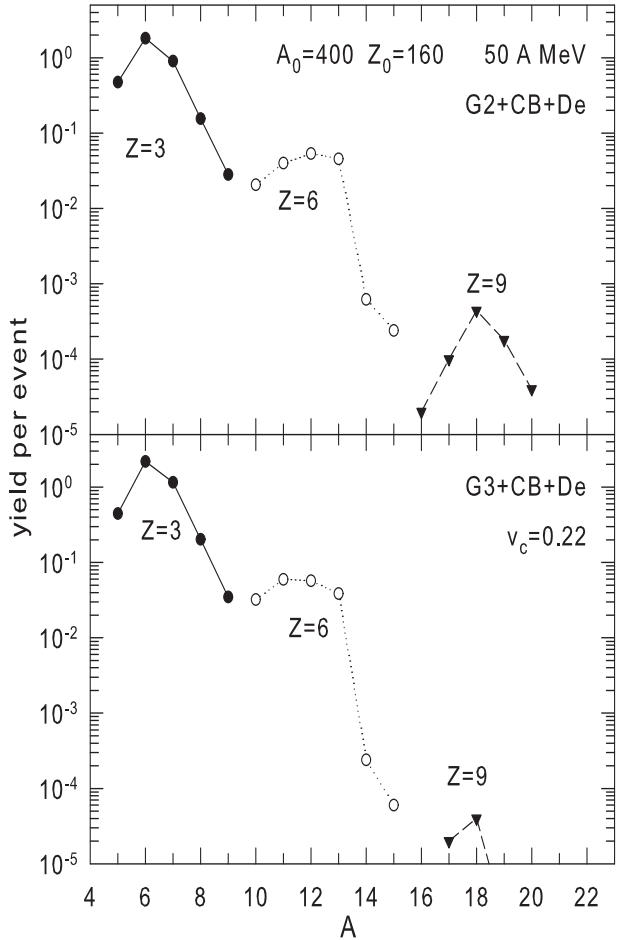}
\caption{\small{ 
Isotope distributions of elements with charges $Z$=3, 6 and 9 after the 
coalescence (CB) and de-excitation (De). The source composition, energy, 
and the coalescent parameter are indicated in the figure. 
The baryon generators G2 (top panel) and G3 (bottom panel) are used. 
}}
\label{fig9}
\end{figure}
\begin{figure}[tbh]
\includegraphics[width=8.5cm,height=13cm]{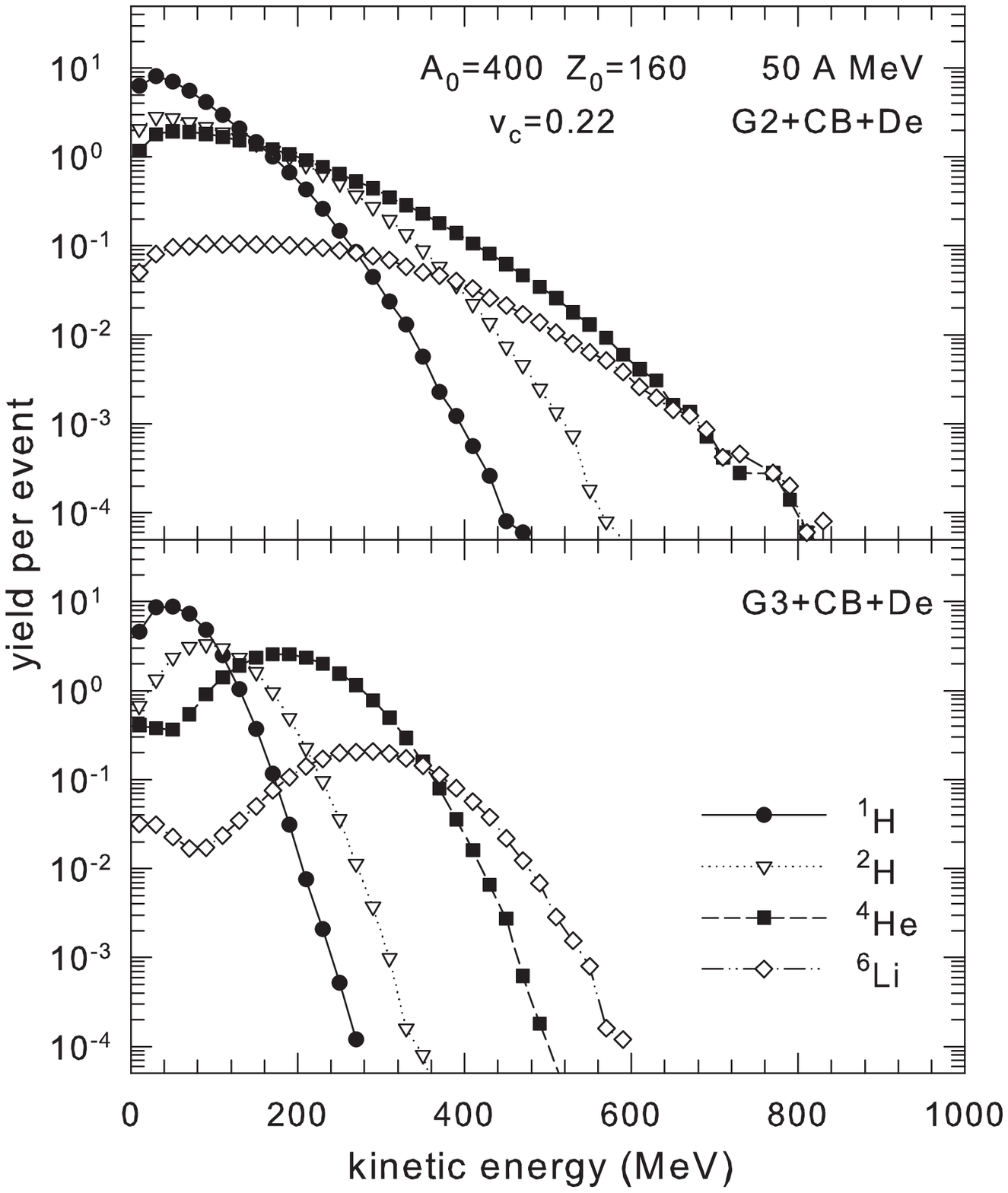}
\caption{\small{ 
Energy distributions of protons and some light particles after the 
coalescence (CB) and the following de-excitation (De). The source composition, 
coalescence parameter, and produced species are indicated in the panels. 
The baryon generators G2 (top panel) and G3 (bottom panel) are used. 
}}
\label{fig10}
\end{figure}

As was done previously in relativistic peripheral collisions and 
heavy-ion collisions at low energies we involve the statistical 
multifragmentation model (SMM) \cite{SMM} to describe the break-up of normal 
nuclear clusters. This approach includes the consistently connected 
multifragmentation, evaporation, fission (for large nuclear systems), and 
Fermi-break-up (for small systems) models. At the same time it reflects 
general	properties of nuclear matter resulting into the phase transition. 
The Fermi-break-up model, which reasonably good describes experimental 
data on disintegration of light nuclei, was generalized also for 
hyper-nuclear systems in Ref.~\cite{lorente}. As well as the evaporation 
and fission models were generalized for hyper-nuclei \cite{Bot16}, and were 
involved for break-up simulations of heavy clusters. Below we demonstrate 
the results obtained after the disintegration of hot primary coalescent 
clusters into the final cold nuclei. 

The secondary de-excitation of primary clusters changes dramatically all 
characteristics of yields and spectra of the nuclei.  
In Figs.~\ref{fig7} and \ref{fig8} we demonstrate 
how the mass distributions of fragments, shown previously in Figs.~\ref{fig2} 
and \ref{fig3}, will change after the de-excitation. For clarity we present 
only few energies $E_0$ and coalescence parameters $v_c$ for the both 
baryon generators. 
It is obviously that the fragment sizes decreases considerably because of 
disintegration of large clusters. In addition, the final fragment 
distributions behave differently than the primary coalescent ones as function 
of the coalescence parameter. For example, the 
increase of $v_c$ is not always leading to the larger fragments: Since 
the excitation is higher then the bigger hot fragments can decay into 
smaller peaces too. 

The isospin content of final fragments (in Fig.~\ref{fig9}) changes also in 
comparison with the primary coalescent clusters (see Fig.~\ref{fig5}). The 
distributions become more narrow and the obtained isotopes concentrate 
closer to the stability line. It is expected since these nuclei have largest 
binding energies. Such a behaviour is typical after the statistical 
disintegration, and it was demonstrated in many previous analysis 
(see, e.g., Ref~\cite{Ogu11}). 

Fig.~\ref{fig10} shows the energy distribution of protons and light fragments 
produced after the de-excitation. In comparison with Fig.~\ref{fig4} one can 
see that many protons with low energy again appear in the system, however, 
as the de-excitation product. By comparing with Fig.~\ref{fig1} we see also that 
high-energy protons can appear in the system (G3 case) as a result of coalescence 
and de-exciation processes. 
The energies of large fragments can be also 
lower because they are the products of the decay of even larger clusters 
which in many cases are composed from the low-energy nucleons. Still the 
flow-like distribution picture with a local maximum remains for G3 
generation. 

\begin{figure}[tbh]
\includegraphics[width=8.5cm,height=13cm]{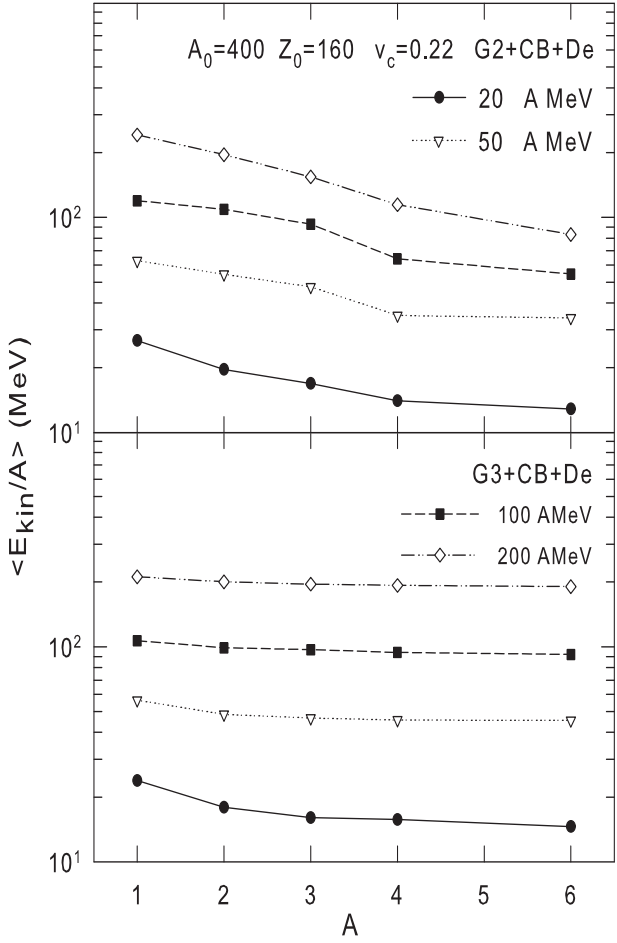}
\caption{\small{ 
Average kinetic energies (per nucleon) of fragments  versus their mass 
number $A$, after the coalescence and the following de-excitation of 
excited clusters. The generators of initial nucleons G2 (top panel) and G3 
(bottom panel) are used. The source composition and energies, coalescence 
parameter, and produced species are indicated in the panels. 
}}
\label{fig11}
\end{figure}

In Fig.~\ref{fig11} we complement this information with the average 
kinetic energy per nucleon of these clusters, at different source energies. 
The top and bottom panels present these energies after G2 and G3 generators 
respectively. This characteristic can be measured in experiments and it is 
often associated with a flow energy. We see some important differences in 
the fragment energies, therefore, it can be used for the identifications 
of the initial dynamical nucleon distributions. In particular, the kinetic 
energy per nucleon is slightly decreasing with mass number $A$ in the 
case of the phase space generation G2. This is because the coalescent large 
fragments are formed predominantly from the slow nucleons which dominate 
after this generation (see Fig.~\ref{fig1}). While after the 
hydrodynamical-like generation G3 the nucleons with a high energy are 
enhanced and uniformly distributed in the space. As a result, after 
the cluster formation and its decay, the fragments from such nucleons have 
approximately the same flow energy per nucleon: It is evident that the 
de-excitation leads to smaller fragments, however, their velocities depend 
on the velocities of the constituent nucleons in the expanded system. 

\begin{figure}[tbh]
\includegraphics[width=8.5cm,height=13cm]{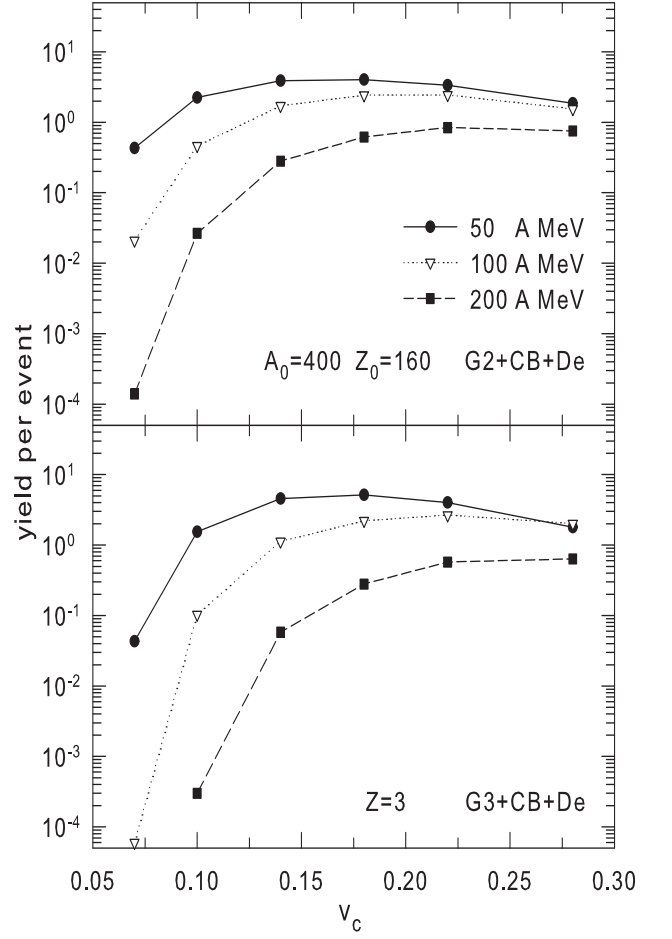}
\caption{\small{ 
Yield of $Z$=3 nuclei as function of the coalescence parameters 
after the coalescence (CB) and the following de-excitation (De). 
The nucleon generators G2 (top panel) and G3 (bottom panel)  are used. The 
source composition and their energies are indicated in the panels. 
}}
\label{fig12}
\end{figure}

It is instructive to show how the yield of final intermediate mass fragments, 
for example, with $Z$=3 can change with the coalescence parameter $v_c$ for 
various source energies. As we mentioned, there is an interplay of two 
effects: The increase of $v_c$ leads to large primary coalescent fragments. 
However, their internal excitation energies are also becoming larger. 
Therefore, as a result of the de-excitation they break-up into smaller final 
nuclear species. One can see from Fig.~\ref{fig12} the yield may have a 
local maximum at some intermediate parameters: There is a trend to increase 
yields of these fragments with $v_c$ in the region of low $v_c$ and to 
decrease the yields at high $v_c$. Also their production decreases for high source 
energies. This is a quite universal behaviour of the fragment production and 
it is manifested for the both generators. 
\begin{figure}[tbh]
\includegraphics[width=8.5cm,height=13cm]{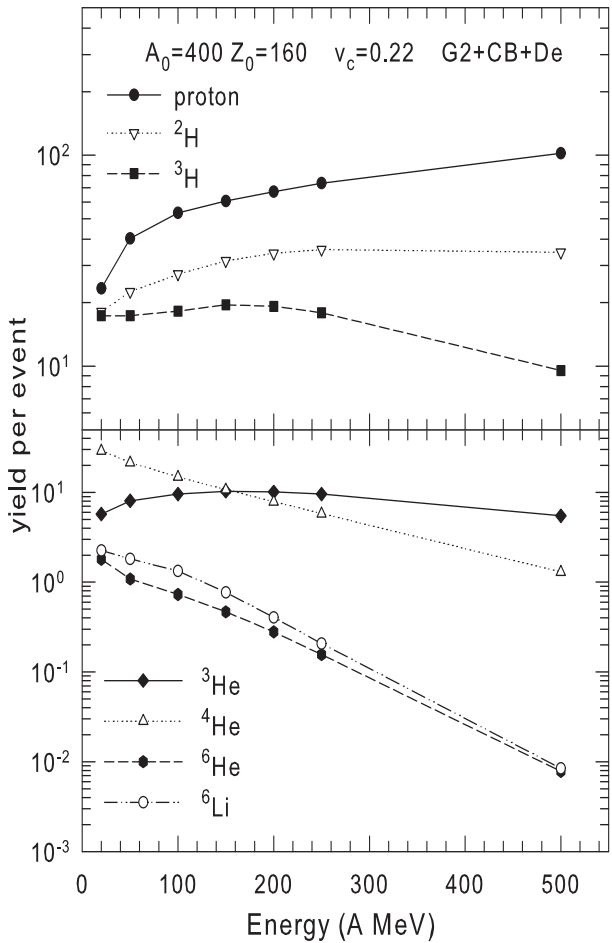}
\caption{\small{ 
The yields of protons and light charged particles after G2 generator and 
the coalescence with de-excitation of hot coalescent clusters (CB+De), 
as function of the source excitation energy. 
The coalescence parameter is $v_c$=0.22$c$. 
The notations for the source and particles are shown in the panels. 
}}
\label{fig13}
\end{figure}
\begin{figure}[tbh]
\includegraphics[width=8.5cm,height=13cm]{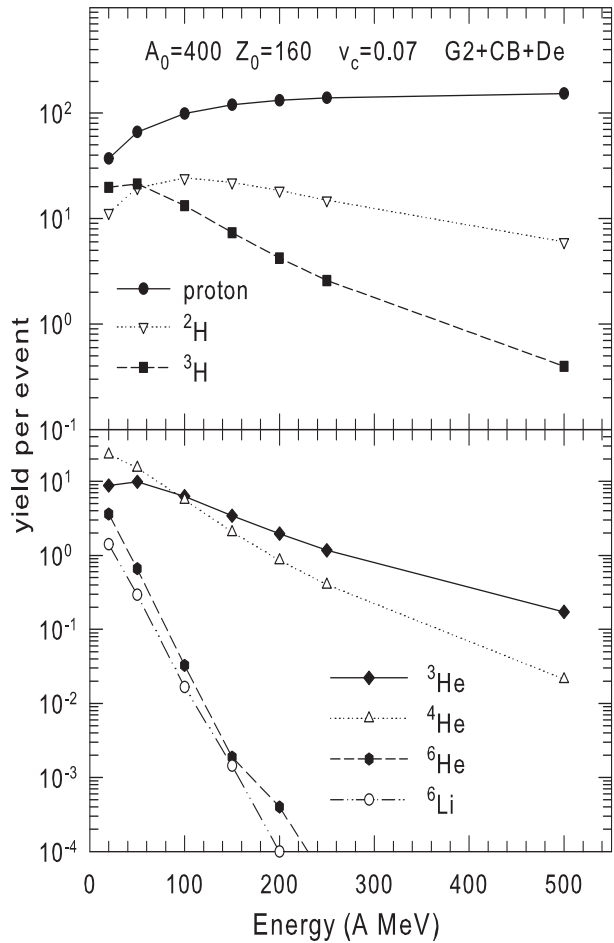}
\caption{\small{ 
The same as in Fig.~\ref{fig13} but for the coalescence parameter 
$v_c$=0.07$c$
}}
\label{fig14}
\end{figure}

The production of lightest charged fragments, such as $p$, $^{2}$H, $^{3}$H, 
$^{3}$He,  and $^{4}$He, dominates in central relativistic heavy-ion collisions. 
Therefore, in Figs.~\ref{fig13} and \ref{fig14} we show how the corresponding 
yields depend on the source excitation energy. For clarity we have selected 
a large $v_c$ which results in big primary clusters with high internal 
excitation energy (Fig.~\ref{fig13}). Also we demonstrate a small $v_c$ 
corresponding to very low excited coalescent clusters (Fig.~\ref{fig14}). 
We show only the results after G2 generator, since using G3 leads to 
qualitatively same conclusions. 
\begin{figure}[tbh]
\includegraphics[width=8.5cm,height=13cm]{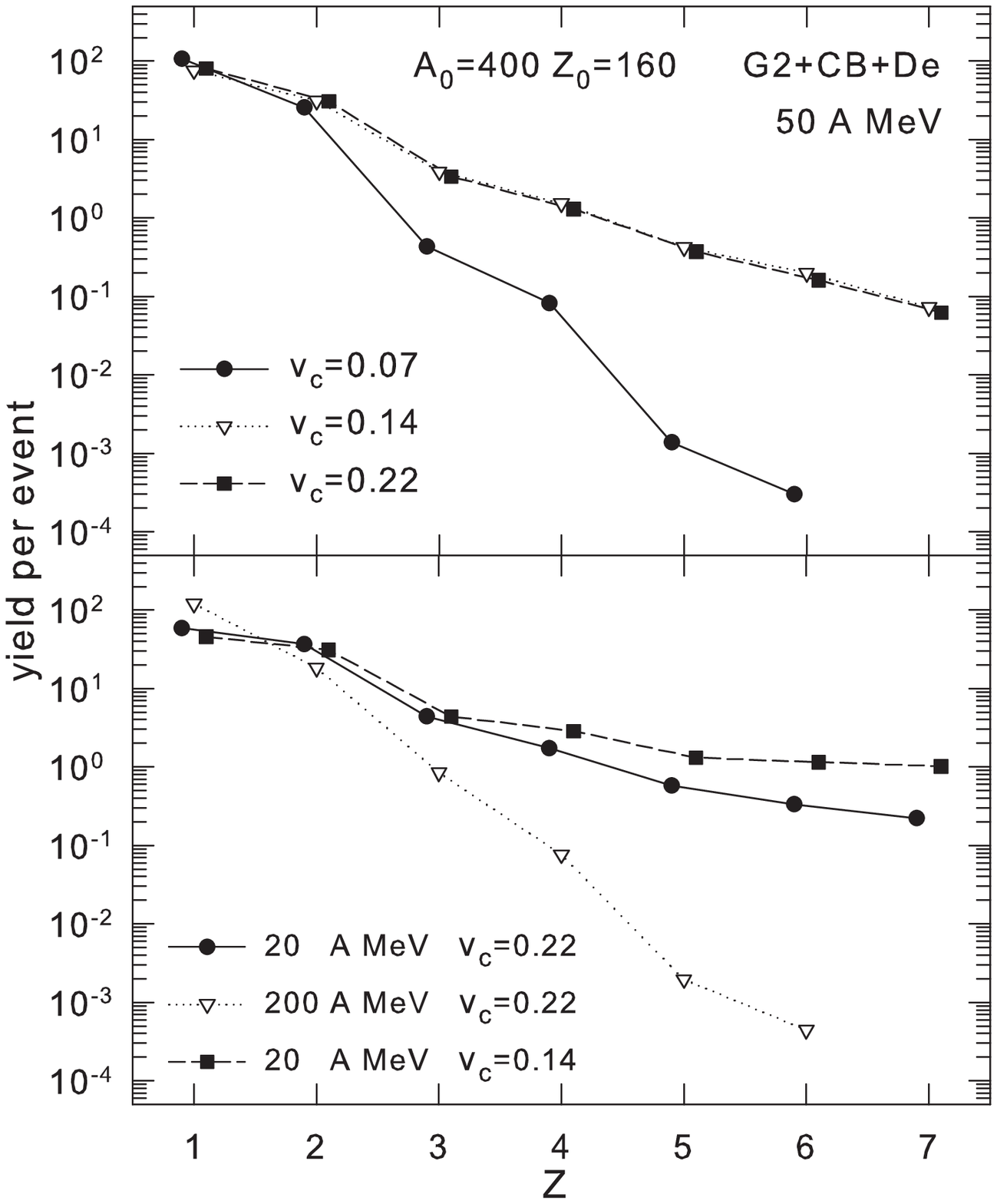}
\caption{\small{ 
Charge yields of light and intermediate mass nuclei after the coalescence 
and de-excitation. The nucleon generator, source composition and energies, 
and coalescence parameters are indicated in the panels. 
}}
\label{fig15}
\end{figure}
\begin{figure}[tbh]
\includegraphics[width=8.5cm,height=13cm]{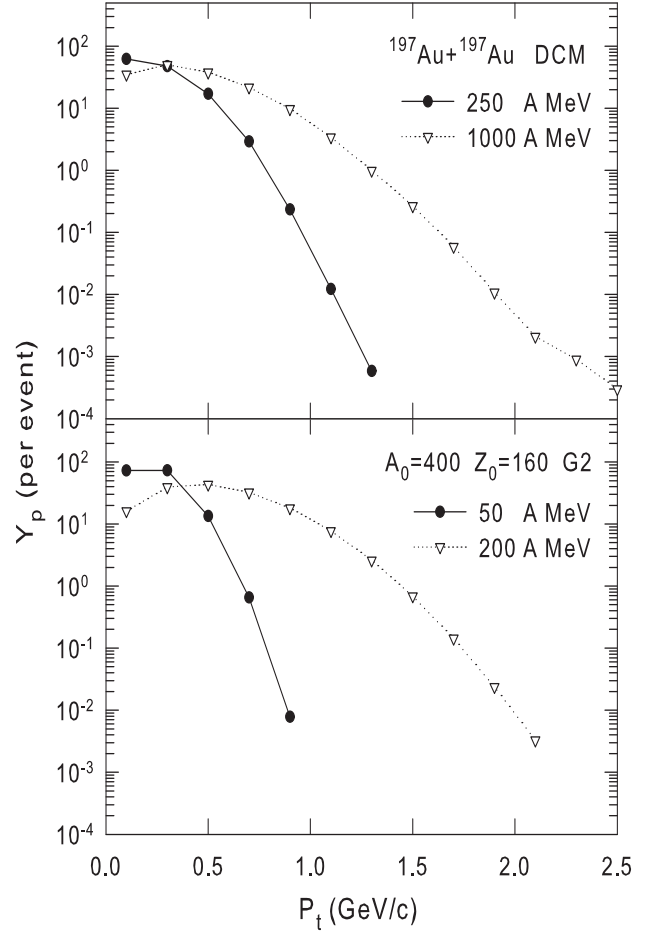}
\caption{\small{ 
Transverse momenta distributions of protons produced after the dynamical 
DCM stage in Au+Au central collisions at energies of 1000 A MeV and 250 A MeV 
(top panel), and after the phase space disintegration of sources (G2) with 
energies of 200 A MeV and 50 A MeV (bottom panel).
}}
\label{fig16}
\end{figure}
\begin{figure}[tbh]
\includegraphics[width=8.5cm,height=13cm]{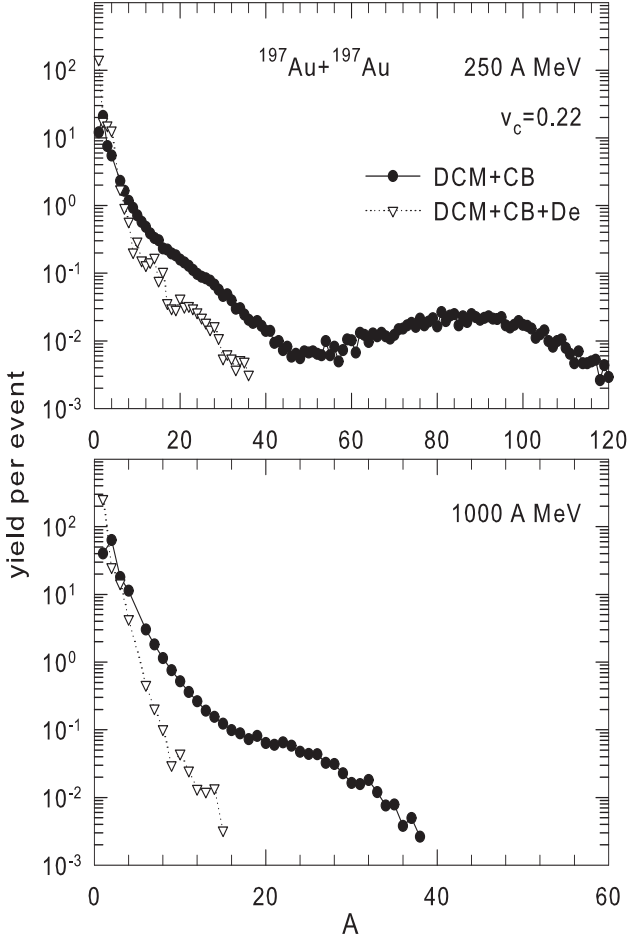}
\caption{\small{ 
Calculated distributions of coalescent clusters (after DCM and coalescence: 
DCM+CB) and final nuclei (after de-excitation: DCM+CB+De) in mass number. 
Beam energies of central collisions of gold nuclei and the coalescence parameter 
are shown in the figure. 
}}
\label{fig17}
\end{figure}
\begin{figure}[tbh]
\includegraphics[width=8.5cm,height=13cm]{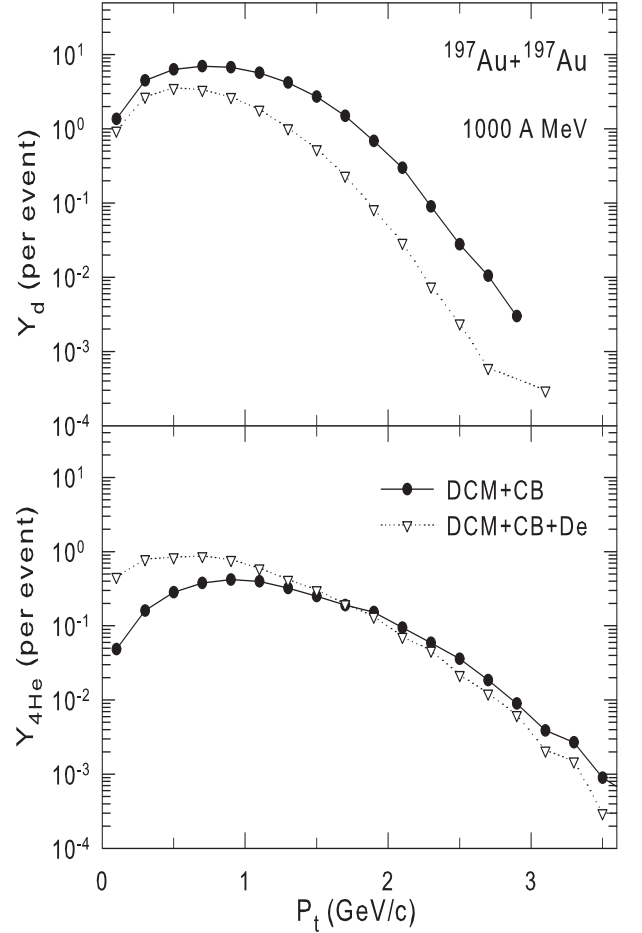}
\caption{\small{ 
Transverse momentum distribution of the coalescence clusters and final 
nuclei of $^2$H (top panel) and $^4$He (bottom panel). In the calculation 
the reaction parameters are as in Fig.~\ref{fig17}. 
}}
\label{fig18}
\end{figure}
\begin{figure}[tbh]
\includegraphics[width=8.5cm,height=13cm]{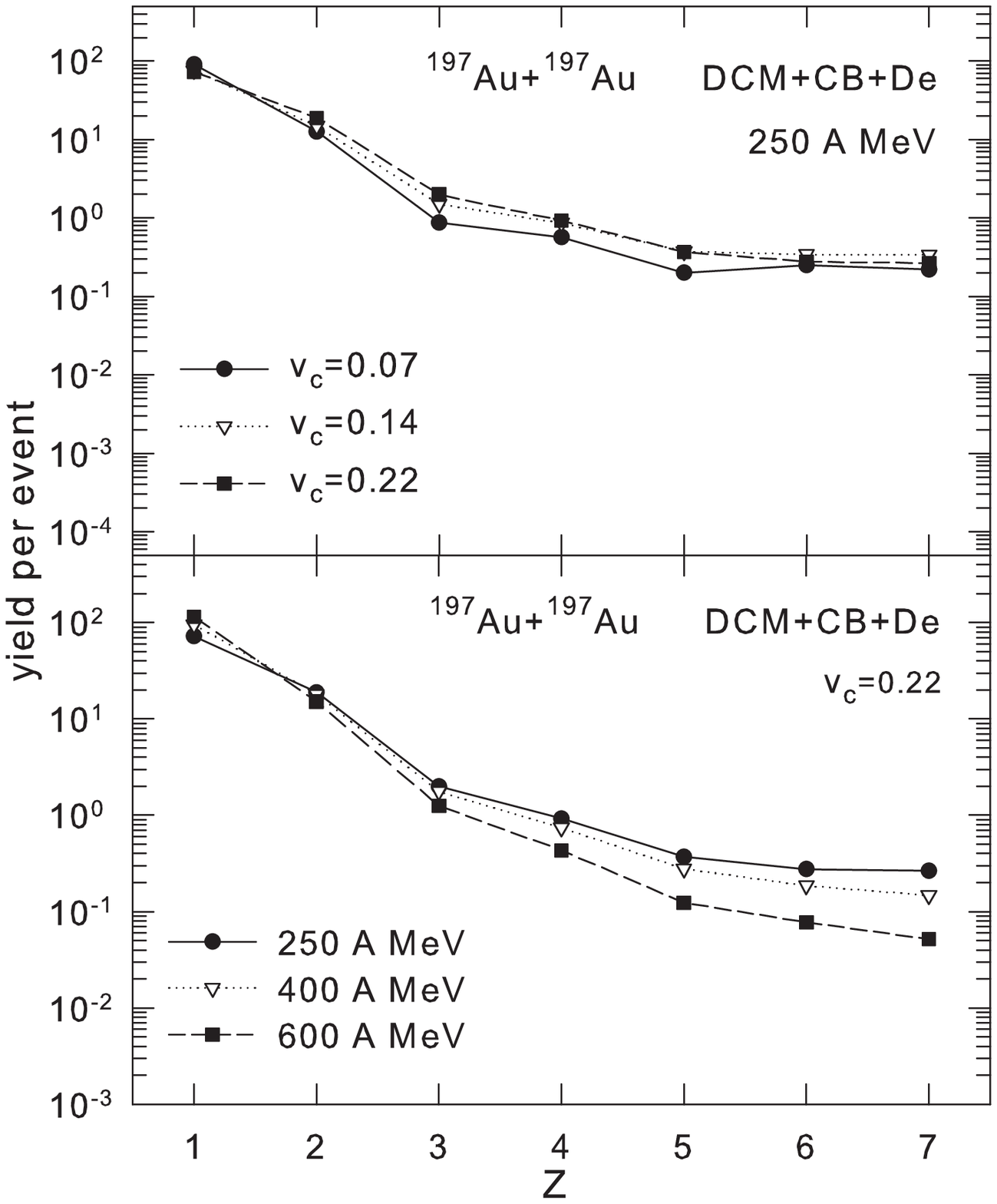}
\caption{\small{ 
Charge yields of light and intermediate mass nuclei in central collisions 
of two gold nuclei obtained after DCM, coalescence, and de-excitation 
calculations. The beam energies and the coalescence parameters are 
indicated in the panels. 
}}
\label{fig19}
\end{figure}

As expected, the yield difference between protons and complex 
particles with $A$=2, 3, 4, and 6 becomes larger at the high source energy, 
since the system disintegrates into smaller pieces. This is an obvious 
consequences of a decrease in production of primary coalescent clusters with 
$A$. However, at relatively low source excitations, when big primary clusters 
are still produced, the situation is different. The internal excitation 
of such clusters is high  and the de-excitation results depend on the 
binding energies of produced species. For this reason the yields of 
$^{4}$He becomes larger than the $^{3}$He yields. This result is quite 
surprising since in the standard coalescence picture (i.e., without 
de-excitation) the yields of large clusters is always lower than the 
small ones. And as one see from the both figures this is true for all 
reasonable coalescence parameters under investigation, though it is more 
pronounced at large $v_c$ when big primary clusters are abundantly produced. 
Another 
interesting result is that the final yield of $^{6}$Li can be larger than 
$^{6}$He in the sources, at big $v_c$. This is also 
related to the slightly larger binding energy in Li. However, this small 
effect is lost at small $v_c$, since the sources are neutron rich and 
the isospin effect dominates by favoring the formation of neutron rich 
nuclei. For this reason the comparison of light cluster yields can help 
to distinguish the internal excitations of primary coalescent clusters 
and find out the production mechanisms in experiments. 

From our experience in multifragmentation reactions, in order to clarify the 
fragment production regularities, it is important 
to look at yields of fragments with $Z\geq$3 too. In Fig.~\ref{fig15} we 
present the charge yields which can be observed in such experiments. We show 
again G2 generator calculations since G3 gives qualitatively similar results. 
As expected for the central collisions of high energy the yield drops with 
$Z$ nearly exponentially. Obviously, the higher energies lead to the smaller 
yields of $Z\geq$3. One can conclude from the analysis of this figure, as well 
as Fig.~\ref{fig12}, that yields of big nuclei can be the largest one at 
intermediate $v_{c}$. Since it provides the best balance between the size of 
primary clusters and their internal excitations leading to the formation of 
intermediate mass nuclei. Actually, such yields are very sensitive 
characteristic for the many-body reaction process, and it is complementary 
to the production of lightest nuclei and protons. Therefore, it should not 
be disregarded in the analysis of experimental data. 

\section{Transport generation of particles and the statistical break-up 
of coalescent clusters} 

Now we consider a practical (and popular) way to treat the relativistic 
ion collisions with the transport models (see, e.g., 
Refs.~\cite{Ton83,urqmd1,urqmd2,HSD,IQMD,giessen}). These models are able to 
describe the initial dynamical stage of the collisions with production 
of many particles including baryons. They are also quite good in description 
of the experimental data. As the first step we have selected the Dubna 
cascade model (DCM) which was since long ago on the market and used for 
analysis of many experiments \cite{Ton83,Bot11,Bot17}. 
Generally, the transport approach should be more realistic one than  the 
simulation of initial nucleons according to the phase space (G2), and 
the hydrodynamical-like flow (G3), since it takes into account explicitly 
the scattering and formation of new baryons. However, if we want to 
analyze experimental data we must take into account the experimental 
filter and make the same selection of the simulated events as in 
the experiment. Sometimes it is difficult to do because of the large 
required statistics. Therefore, the G2 and G3 simulations could be very 
useful in order to find the correct way for extracting physical information 
from the data. 

In Fig.~\ref{fig16} we show the proton transverse momenta predicted by the DCM 
in the case of central (the impact parameter is less than 3 fm) collisions of 
Au on Au at energies of 250 and 1000 A MeV in the laboratory system. For 
qualitative comparison with our analysis in the previous section we show also 
the corresponding results obtained with G2 generator for the sources with 
energies of 50 and 200 A MeV for the $A_0$=400 and $Z_0$=160 system.  These 
source energies are only slightly lower than the corresponding center-of-mass 
energies of the colliding nuclei. One can see 
some differences which should influence the following coalescence process. 
For example, DCM produce more protons with very low transverse momenta and 
the distributions are more broad. 

In the DCM case the same procedure was taken for the coalescence (CB) and 
de-excitation (SMM) of hot coalescence fragments. However, in the 
Monte-Carlo DCM code the primary nucleons and hyperons can be produced 
in different time moments during the whole cascade stage which lasts for 
10--30~fm/c. Therefore, in addition to the relative velocity coalescence 
criterion we suggest that the baryons should be close in the coordinate 
space when this dynamical cascade stage ends. 
In particular, all baryons consisting of a cluster with mass number $A$ 
should be inside the sphere with radius of $R=R_0 A^{1/3}$ from the center 
of mass of the cluster. We take $R_0$=2~fm, as it is obtained by extracting 
the freeze-out volume information in the multifragmentation experiments 
(see Refs.~\cite{Vio01,Pie02,FASA}), and approximately corresponds to 
$\rho_c$ density suggested for G2 and G3 generations (see Section 3). 

We present in Fig.~\ref{fig17} the mass distributions of nuclear clusters 
produced after the coalescence of the cascade nucleons (DCM+CB) and after 
their following de-excitation (DCM+CB+De) into cold nuclei. As previously 
we use the SMM model for the de-excitation description. The regularities 
are similar to the ones demonstrated previously for the excited sources in 
Figs.~\ref{fig2}, \ref{fig3}, \ref{fig7}, and \ref{fig8}. We expect that the 
quite big nuclei will be observed in experiments at the collision energies 
around 250 A MeV, and there is an essential decrease of their yields 
with increasing energy up to 1 A GeV and to higher energies. 

For the same reaction the transverse momentum distributions 
for $^2$H and $^4$He nuclei after both the coalescence and the cluster 
de-excitation are shown in Fig.~\ref{fig18}. The de-excitation leads to the 
essential production of these nuclei with lower momenta, and to a more steep 
decrease of spectra with $p_t$. This trend is more pronounced for large 
nuclei, therefore, such nuclei should predominantly have low transverse 
momenta. 

By using the full DCM+CB+De  approach we demonstrate in 
Fig.~\ref{fig19} how the charge distributions of the final nuclei 
modify depending on the coalescence parameter $v_c$ and the beam 
energy. As one can see we have qualitatively the same evolution  
as was shown in Fig.~\ref{fig15}. However the different initial nucleon 
distributions 
lead to slightly different results: In the DCM case of the 250 A MeV beam 
energy the yield of intermediate mass fragments changes very weak with 
the coalescence parameter. The vicinity of the generated nucleons in 
the velocity and coordinate space after DCM gives a chance to produce 
relatively big primary coalescence clusters even at small $v_c$. In the 
same time the low excitation energy allows for surviving big fragments 
after de-excitation. At large $v_c$ the considerably larger clusters 
are produced. However, they are more excited and can decay into small 
fragments approximately of the same size as at the smaller $v_c$. 
By increasing beam energies this effect disappears and we obtain an 
expected regular decrease of the charge yields. 

\begin{figure}[tbh]
\includegraphics[width=8.5cm,height=13cm]{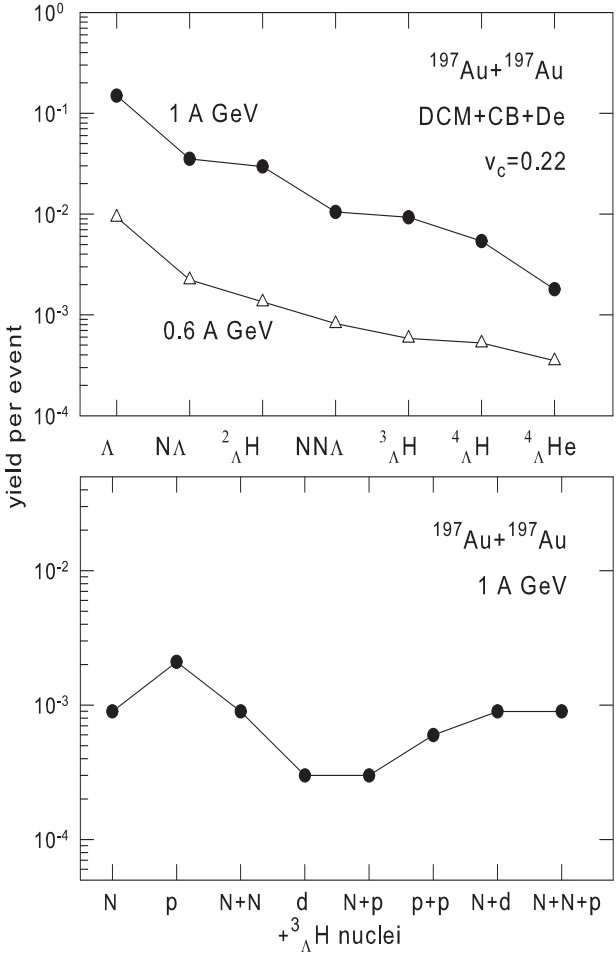}
\caption{\small{ 
Yields of hypernuclei produced in central collisions 
of two gold nuclei after DCM, coalescence, and de-excitation 
calculations. Top panel presents the full yields per event. 
The yields of correlated particles (neutrons, proton, deutrons) in 
channels with the $^3_\Lambda$H production are in the bottom panel. 
The beam energies are indicated in the panels. 
}}
\label{fig20}
\end{figure}

The DCM can simulate the production of new baryons, e.g., hyperons. 
Previously a good comparison of DCM 
with the strangeness production was shown in Refs.~\cite{Bot11,Bot17}. 
Therefore, the 
predictions for hypernuclei can be given within same coalescence and 
statistical de-excitation mechanisms. The 
coalescence and statistical models were generalized for hyperfragments in 
Refs.~\cite{Bot07,Buy13,Bot16,Bot15,lorente}. 
By using this approach we shown the yields of light hypernuclei in 
Fig.~\ref{fig20}. We demonstrate some results for the central collisions at 
the beam energies of 600 A MeV and 1 A GeV which are below the threshold for 
the hyperon formation in nucleon-nucleon interaction. At these energies the 
$\Lambda$ hyperons are produced because of the secondary interactions at the 
cascade stage. Actually, the hypernuclei from such subthreshold 
processes are very important since 
their productions depends on subtle details of hyperon-nucleon interactions. 
Besides well known hydrogen and helium hypernuclei in the top panel we 
show predictions for exotic neutron-$\Lambda$ (N$\Lambda$), proton-$\Lambda$ 
($^{2}_\Lambda$H) , and 
neutron-neutron-$\Lambda$ (NN$\Lambda$) hypernuclei, which are discussed in 
the literature \cite{alice,HypHI} in order to facilitate their experimental 
searching. 

We believe it would be instructive to justify in experiment directly the 
secondary de-excitation of primary coalescent hyper-clusters. For this 
purpose in the bottom panel of Fig.~\ref{fig20} we demonstrate the decay 
channels leading to the production of $^3_\Lambda$H  nuclei for the 
1 A GeV energy case. The charged particles can be easily detected in modern 
experiments and this correlation measurement would be important 
confirmation of the reaction mechanism. 

\begin{figure}[tbh]
\includegraphics[width=8.5cm,height=13cm]{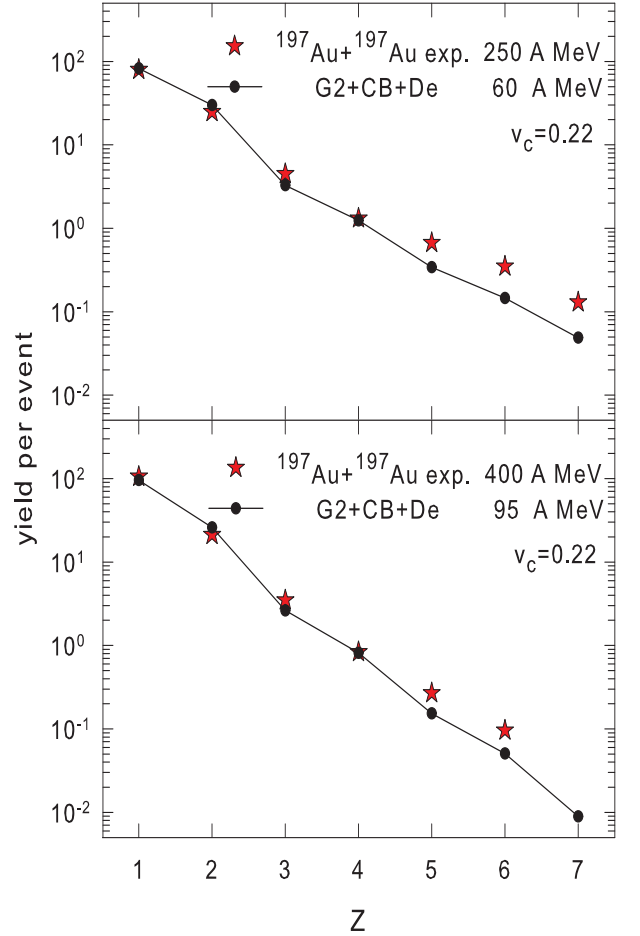}
\caption{\small{ 
Yields of nuclei versus their charge $Z$. The red stars are the FOPI 
experiment \cite{FOPI1997,FOPI2010}. The parameters for the 
calculations including the nucleon generation in Au+Au source, 
coalescence and statistical 
de-excitation are shown in the panels. 
}} 
\label{fig21}
\end{figure}
\begin{figure}[tbh]
\includegraphics[width=8.5cm,height=13cm]{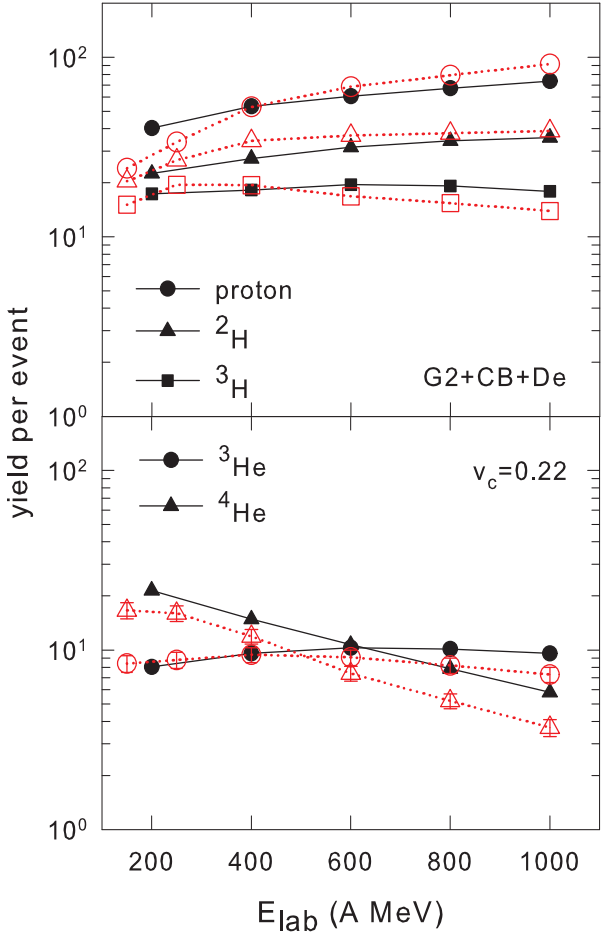}
\caption{\small{ 
Yields of lightest nuclei as function of the beam energy in Au+Au collisions. 
Red symbols connected with the dashed lines are FOPI experimental data 
\cite{FOPI2010}. Black symbols connected with solid lines are our (G2+CB+De) 
calculations with the corresponding center of mass energies for Au+Au sources. 
}}
\label{fig22}
\end{figure}
\begin{figure}[tbh]
\includegraphics[width=8.5cm,height=13cm]{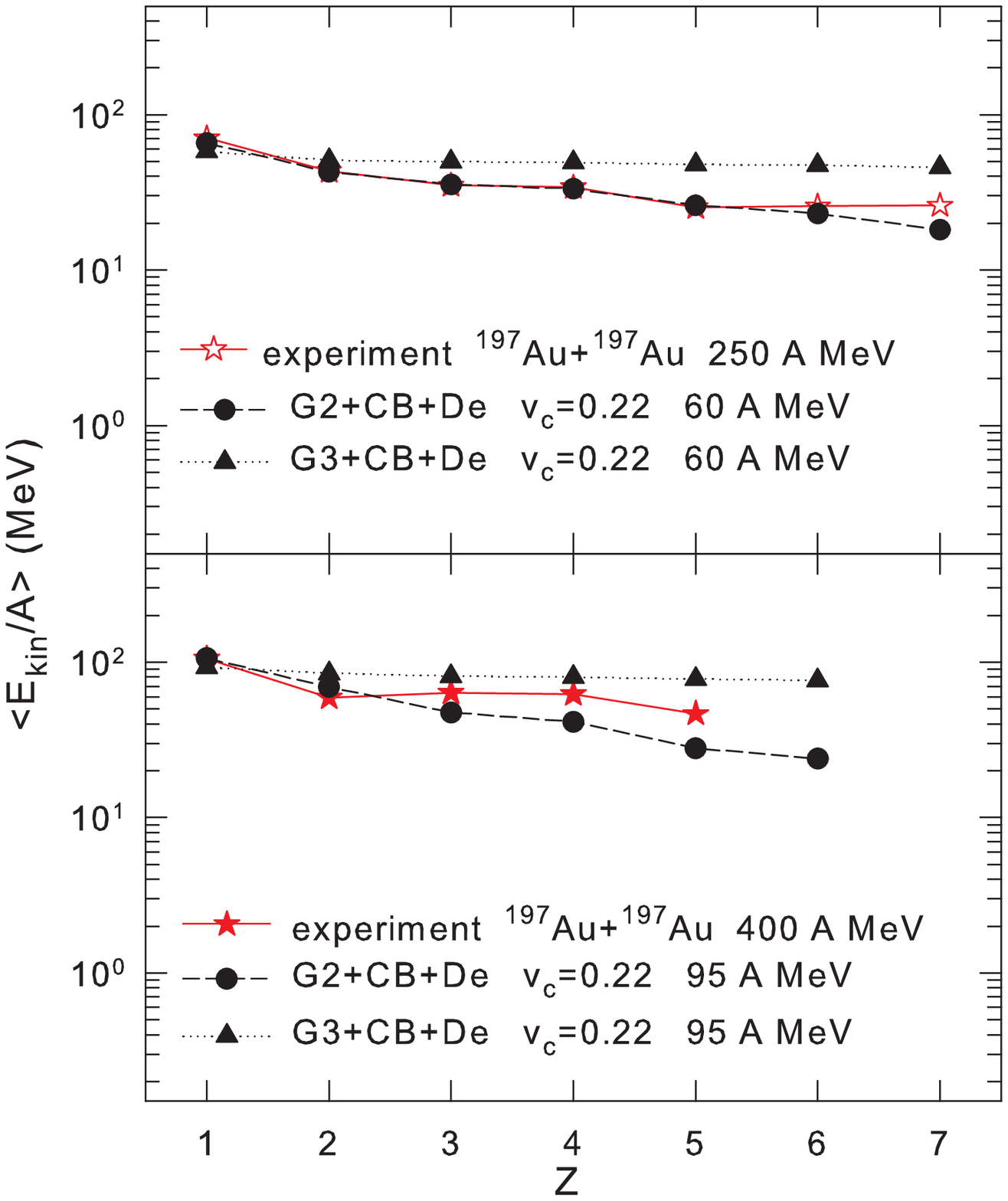}
\caption{\small{ 
Mean kinetic energy (per nucleon) of charged nuclei for central 250 A MeV 
(top panel) and 400 A MeV (bottom panel) of Au+Au collisions. Experimental data are 
in red color \cite{FOPI2010}. The parameters for our calculations (as in 
Fig.~\ref{fig21}) are noted in the panels. 
}}
\label{fig23}
\end{figure}

\section{Analysis of FOPI experimental data} 

In order to confirm the proposed mechanisms we should analyse experimental 
data. We have selected 
the FOPI data on light nuclei produced in central Au+Au collisions, since 
they are the most full and systematic ones in the present time 
\cite{FOPI2010}. 
There were attempts to analyze it with the coalescence and statistical 
prescriptions \cite{Neu03,Som19}. 
We should note that these data are obtained with the selection of the central 
collisions with the ERAT criterion \cite{FOPI1997}, 
which suggests a considerable isotropy 
of the produced particles in the center of mass system. This isotropy is 
naturally provided by the G2 and G3 generations. We have found from 
the momentum analysis that the DCM calculations are not able to provide 
such an isotropy for central events. Because DCM predicts much more 
nucleons with low transverse and, respectively, high longitude momenta 
(see, for example, the comparison presented in Fig.~\ref{fig16}). 
In principle, the DCM sample can be improved by drastic 
increasing the statistics and by the special selecting the central events 
which fulfil the ERAT criterion. That would require much more calculations. 
However, the goal of our present analysis is to show the consistency 
of our approach to the observations. We believe that the simple assumptions 
existing in G2 and G3 generations are sufficient for it. 

In Fig.~\ref{fig21} we demonstrate the integrated charge yields extracted 
in the FOPI experiment with our calculations including G2 generation. We 
consider this kind of generation as the most adequate one for this case, 
since it provides very broad nucleon momentum distributions, as it could be 
expected in the case of realistic transport approaches (Fig.~\ref{fig16}). 
For this analysis we have taken $v_c$=0.22 $c$ which gives moderate internal 
excitations of the primary clusters. As we've noted previously, this velocity 
is of the same order as the Fermi velocity inside nuclei, and provides the 
excitation energies around the nucleus binding energy ($\sim 10$ MeV per 
nucleon, see Fig.~\ref{fig6}). 
The total system was taken as having 394 nucleons with 158 protons 
(Au+Au system). Since the G2 generation is determined by the 
excitation energy we have taken the center of mass energies 
corresponding to the beam colliding energies. The results for the 
energies of 250 A~MeV and 400 A~MeV, which correspond to the center of 
mass energies 60 MeV and 95 MeV per nucleon are shown. For more high beam 
energies 
the nuclei with $Z\goo$3 were practically not observed in the experment. One can see a quite 
reasonable agreement with the data in this case. 
We believe, however, the involvement of other observables is necessary 
to get the consistent theory description. 

The instructive information can be obtained by analysing the lightest 
particle yields, as was shown in Figs.~\ref{fig13} and \ref{fig14}. 
Fig.~\ref{fig22} presents yields of  $p$, $^{2}$H, $^{3}$H, $^{3}$He, 
and $^{4}$He versus the beam energy. 
Within our approach we reproduce the behaviour of their production as 
function of the energy. A very interesting experimental feature has no a 
reasonable explanation up to now: There is the cross-over of the $^{3}$He 
and $^{4}$He yields.  
At low beam energies $^{4}$He dominates, while at high energy we have 
the standard 'coalescence' situation when $^{3}$He is more produced than 
$^{4}$He. As we have pointed above (Section 4), 
the enhanced yield of $^{4}$He at low energies can 
be naturally explained as a result of the secondary de-excitation of primary 
coalescent clusters. $^{4}$He formation is dominating during the statistical 
processes because the binding energy of $^{4}$He is essentially larger 
than $^{3}$He. On the other hand at very high energy the primary coalescent 
clusters becomes rather small, therefore, the nuclei of smaller sizes 
have more chances to be produced. There were no attempts to explain this 
experimental cross-over of helium with previous theories. 

Fig.~\ref{fig23} shows the average kinetic energies of the produced nuclei. 
Our calculations with G2 and G3 generations do also reproduce it reasonably 
well. As obvious from the initial nucleon energy distributions 
(Fig.~\ref{fig1}) G3 provides higher average energies for big nuclei 
because a lot of nucleons have high initial velocity, as a consequence a 
regular flow profile. 
However, in order to conclude on the nature of the 
flow we should look also at the full energy distributions 
(see Figs.~\ref{fig4} and \ref{fig10}), if they are 
available in experiment. We hope, in future, we will get such data. 
The result may also depend on the selection of experimental events, and this 
can be taken into account within our approach. 

As seen from the analysis of the experimental data  we have obtained the 
qualitative agreement for the main observables, i.e., for the nuclei yields, 
their beam energy dependence,  
and their energy spectra. It was not possible to rich consistently in 
previous statistical or dynamical analyses of these data. 
Our consistent description of all characteristics can be considered as the 
confirmation of the 
mechanism suggested for the production of complex nuclei in central 
collisions. 

\section{Conclusion}

During last decades there is permanent increasing the number of experiments 
measuring nuclear reactions in central relativistic nuclear collisions. The 
yield of light nuclei is one of the essential observable. There is a 
reasonable assumption that this yield should be described by transport 
dynamical models if we include a relevant baryon/nucleon interactions 
at low energies. However, the modern transport approaches are designed mainly 
for the description of high energy interactions, including both ions and 
hadrons ones. 
A sophisticated low energy nucleon interaction and other theory ingredients 
important for the realistic description of nuclei (e.g., the calculations 
of real wave functions, antisymmetrization, many-body forces, and so on) are 
usually 
beyond this scope because of complexity of this many body problem. For this 
reason a phenomenological coalescence approach is often used to describe the 
nuclei yields by assuming that the baryons are combined in the final state. 
This simple phenomenology disregards many aspects of low-energy collective 
interactions and may lead to wrong conclusions on the clusterization nature. 
In present our study we try to overcome the problem by considering neighbour 
baryons produced after the dynamical stage as clusters at certain subnuclear 
densities where the baryons are still interacting. This interaction can lead 
to the nuclei production and can be described in the statistical way. 

In our approach, as the first step, after generating the initial baryons and 
their momenta, we involve the generalized coalescence model (CB) which forms 
big excited coalescent-like clusters. In such clusters the baryons 
move respect each other and interact by producing final nuclei. As we 
know the statistical description of such processes is commonly accepted 
for many physical phenomena, for example, in multifragmentation 
\cite{SMM,MMMC}. A crucial 
question is the excitation energy of such primary clusters. Namely it 
determines if the finite system can be considered as an equilibrated one 
during the reaction. At low internal excitations the clusters' baryons are 
together during a long time, therefore, the thermalized conditions are 
fulfilled. On the contrary, at very high excitations the baryons should fly 
away fastly and the equilibrium criterion can be violated. Remarkably, 
however, that in order to describe the experimental data we should take the 
excitation energy of around 10 MeV per nucleon. This is close to the nuclear 
binding energy, and it is similar to the energy which we have previously 
extracted from 
the analysis of the projectile/target residue multifragmentation. This fact 
may tell us that there are common conditions for establishing equilibrium in 
finite nuclear systems. 

We note that previously only one equilibrated excited nuclear source was 
assumed in the statistical models' applications for the description of 
nuclei production in central nucleus collisions (see, e.g. 
Refs.~\cite{Neu03,SMM}). It might be formed as a 
results of the full or partial fusion of the colliding nuclei. Our approach 
with many such sources (i.e., excited coalescence clusters) allows for 
more consistent description of the reaction. The flow of the produced 
particles can be here explained as a dynamical motion of the clusters. The 
nuclei of small sizes which dominate at the very high collision energy can 
be now explained not as a result of very high temperatures of one source 
but also as a result of decreasing primary sizes of the coalescent clusters. 
In addition, 
we obtain a new physical constraint on the excitation energies of equilibrated 
nuclear sources of finite sizes in fastly expanded big systems. 

We have theoretically investigated the main regularities of such nuclei 
production, in particular, charge and isotope yields, their kinetic energies. 
We have also investigated the influence of the initial baryon generating 
stage, which can be described by dynamical models. We can reasonably describe 
the recent FOPI data, however, we need new experimental data for verifying 
this approach. In addition to inclusive yields and energy spectra, 
the particle 
correlations and the correlated yields, which come after decay of primary hot 
clusters, should be the adequate observable. This mechanism allows for a new 
interpretation of the baryons and nuclei yields, since the secondary 
'statistical' interaction can change the baryon characteristics from the 
primary 'dynamical' ones. As we found it is certainly expected in the 
collisions with the beam energy less than 1 GeV per nucleon. However, the 
higher energies are more interesting since they lead to the production 
of new particles and nuclei, e.g., hypernuclei. In this case we can obtain 
novel information on hyperons interaction at low energy in matter and new 
exotic species. Such kind of research can be possible at the new generation of 
ion accelerators of intermediate energies, as FAIR (Darmstadt), NICA (Dubna), 
and others. It is promising that new advanced experimental installations for 
the fragment detection will be available soon \cite{aumann,frs}. 

\begin{acknowledgments}
The authors thank our colleagues for stimulating discussions and support 
of this study: J. Pochodzalla, W. Trautmann, Ch. Scheidenberger, Y. Leifels, 
A. Le-Fevre, R. Ogul, L. Bravina.  A. S. Botvina acknowledges the support 
of BMBF (Germany). N. B. acknowledges the Scientific
and Technological Research Council of Turkey (TUBITAK)
support under Project No. 118F111. M. B. and N. B. acknowledges that the 
work has been performed in the framework of COST Action CA15213 THOR. 
The authors thank the Frankfurt Institute for Advanced Studies (FIAS), 
J. W. Goethe University, for hospitality. 
\end{acknowledgments}

\end{document}